\documentclass[journal=jacsat,manuscript=article,maxauthors=7,
etalmode=truncate]{achemso}
\setkeys{acs}{maxauthors=7}
\setkeys{acs}{etalmode=truncate}

\usepackage{amsmath}
\usepackage{amsmath,amssymb}
\usepackage{graphicx}
\usepackage{hyperref} 
\usepackage{caption}
\usepackage{color}
\usepackage{dcolumn}
\usepackage{bm}
\usepackage{float}
\usepackage{subfig}
\usepackage{soul}
\usepackage[normalem]{ulem}
\usepackage{microtype}
\usepackage[version=3]{mhchem} % Formula subscripts using \ce{}
\usepackage{booktabs}
\usepackage{multirow}
\usepackage{booktabs}
\usepackage{longtable}
\usepackage[table]{xcolor} % for \cellcolor
\usepackage{xr}

\makeatletter
\newcommand*{\addFileDependency}[1]{%
  \typeout{(#1)}%
  \@addtofilelist{#1}%
  \IfFileExists{#1}{}{\typeout{No file #1.}}%
}
\makeatother

\newcommand*{\myexternaldocument}[1]{%
  \externaldocument{#1}%
  \addFileDependency{#1.tex}%
  \addFileDependency{#1.aux}%
}

\myexternaldocument{si}

\usepackage{rotating}

\SectionNumbersOn
\usepackage[dvipsnames]{xcolor}

\author{Raghunathan Ramakrishnan} \affiliation[TIFR Hyderabad]
{Tata Institute of Fundamental Research, Hyderabad 500046, India}  \email{ramakrishnan@tifrh.res.in}

\title{
A Chemical Space Perspective on Diastereomeric Barriers in Alkylperoxy-to-Hydroperoxyalkyl Isomerization
}

\begin{document}

\begin{tocentry}
\centering 
  \includegraphics[width=6.7cm]{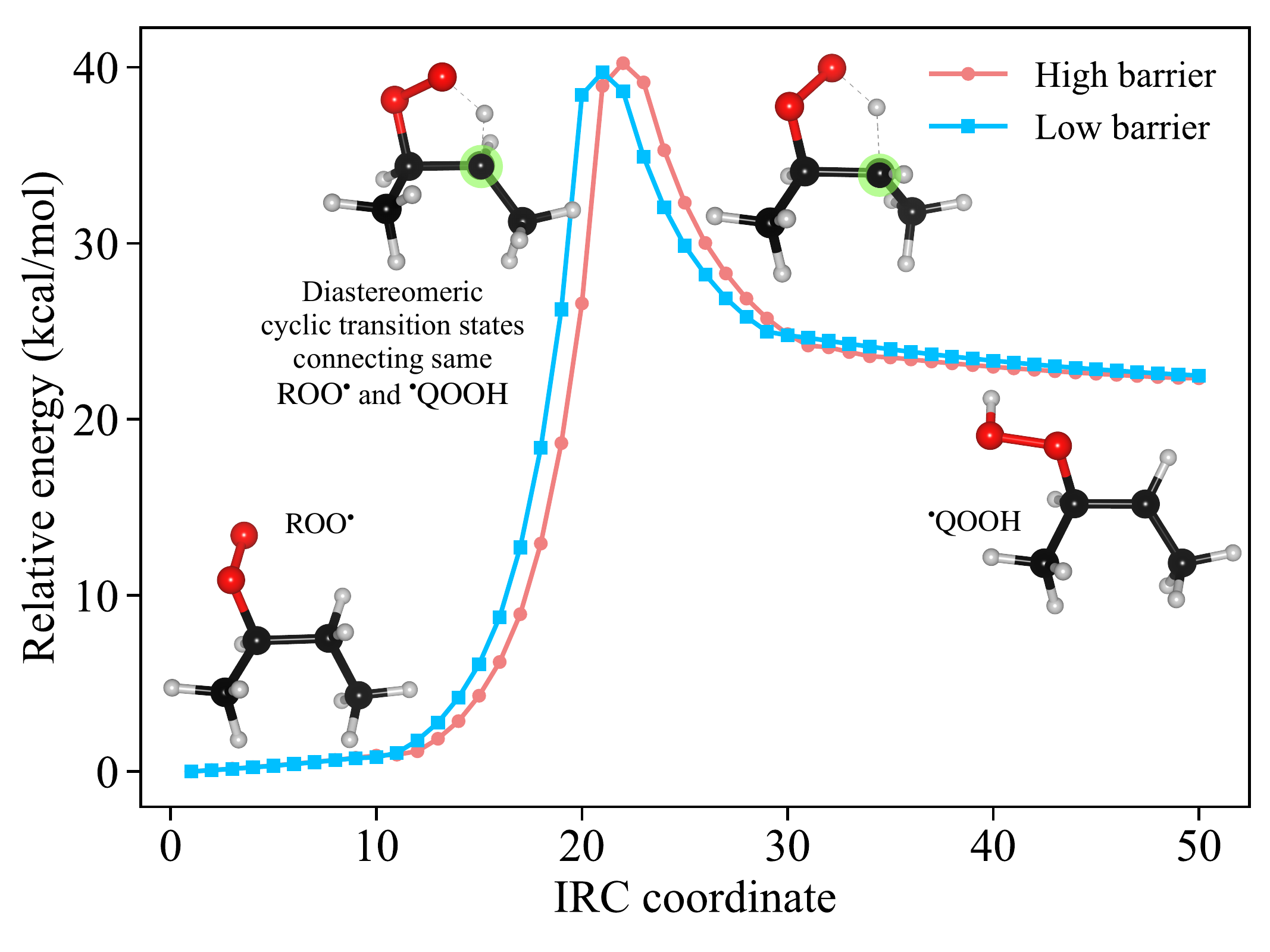}
\end{tocentry}

\begin{abstract}
Low-temperature hydrocarbon autooxidation involves radical intermediates whose reactivity depends not only on the stereochemistry of the intermediates themselves, but also on that of the transient species encountered along the reaction path. 
This study offers large-scale evidence for the importance of stereochemistry in low-temperature autooxidation by propagating stereochemical information from 498 \({\rm C}_1\)–\({\rm C}_7\) hydrocarbons through radical formation, \(\mathrm{O_2}\) addition, and the isomerization of alkylperoxy (\(\mathrm{ROO^\bullet}\)) radicals to hydroperoxyalkyl (\(\mathrm{{}^\bullet QOOH}\)) radicals. 
The resulting dataset comprises density-functional-theory-level data for 5,356 species, including 2,324 cyclic diastereomeric transition states associated with 1,162 unique \(\mathrm{ROO^\bullet \rightarrow {}^\bullet QOOH}\) isomerization reactions, with transition-state connectivity confirmed by intrinsic reaction coordinate analysis. 
Explicit stereochemical treatment reveals that diastereomeric pathways may be either degenerate or separated by more than 60 kcal/mol, with the magnitude of these differences governed by steric strain at the carbon bearing the peroxyl group. 
These results show that constitutionally collapsed molecular representations can systematically miss kinetically relevant reactive channels and provide a foundation for stereochemistry-aware mechanism generation, rate estimation, and predictive combustion modeling.
\end{abstract}


\section{Introduction}

Hydrocarbon combustion proceeds through complex radical-chain reaction networks involving multiple competing elementary steps over a wide temperature range\cite{norrish1935theory,franklin1967mechanisms,hucknall2012chemistry,bugler2015revisiting,kecceli2019automated,westbrook2000chemical}. 
These highly exothermic processes involve chain-initiation, propagation, and termination chemistry\cite{mayo1968free,orlando2012laboratory}. 
In low-temperature oxidation, alkyl radicals (R$^\bullet$) rapidly add O$_2$ to form alkyl peroxy radicals (ROO$^\bullet$), which can either continue chain propagation or undergo intramolecular H-transfer to form $^\bullet$QOOH radicals, thereby enabling subsequent O$_2$ addition and chain-branching pathways central to autooxidation\cite{zador2011kinetics}. 
The kinetics of these reactions depend systematically on molecular structure\cite{miyoshi2011systematic}. 
Beyond combustion, peroxy-radical H-shift reactions also play an important role in atmospheric oxidation networks\cite{moller2019importance}. 
Since $^\bullet$QOOH radicals are highly reactive and short-lived, their direct spectroscopic detection has long been challenging, and they have often been identified only indirectly through their dissociation products\cite{qian2024infrared}. 
Accordingly, only a limited number of such combustion- and atmosphere-relevant intermediates have been observed experimentally\cite{welz2012direct,taatjes2013direct,su2013infrared,savee2015direct,hansen2021watching,hansen2022infrared}.

Stereochemical effects can be kinetically important in low-temperature hydrocarbon oxidation, especially in the \(\mathrm{ROO^\bullet \rightarrow {}^\bullet QOOH}\) isomerization step that controls access to downstream chain-branching chemistry\cite{zador2011kinetics,rotavera2021influence}. 
This situation is distinct from symmetry-equivalent reactive arrangements, for which the observed rate is accounted for through an appropriate reaction-path degeneracy or symmetry-number correction\cite{schlag1963symmetry,fernandez2007symmetry,pollak1978symmetry}. 
In some cases, a single \(\mathrm{ROO^\bullet}\) intermediate can access the same \({}^\bullet\mathrm{QOOH}\) potential well through two distinct isomerization pathways that pass through unique transition states differing in three-dimensional arrangements of atoms. 
Such transition-state pairs are diastereomeric when the H-shift generates stereochemically distinct saddle-point structures, particularly when the \(\mathrm{ROO^\bullet}\) intermediate is already chiral. 
Since such pathways are not related by reflection or any symmetry operation, they must be treated as distinct reactive channels rather than as symmetry-equivalent or permutationally equivalent pathways. 
These species have therefore been termed ``ephemeral diastereomers,'' emphasizing that the stereochemical distinction is expressed only at the transition state rather than as an isolable intermediate\cite{danilack2021diastereomers}. 
For example, from a single conformer of \(\mathrm{ROO^\bullet}\), H abstraction from opposite faces of a secondary C--H bond can lead to two diastereomeric transition states. 
These pathways produce the same \(\mathrm{{}^\bullet QOOH}\) radical connectivity but may differ in the conformation of the OOH group relative to the newly formed ring at the transition state\cite{copan2024radical}. 
The resulting \(\mathrm{{}^\bullet QOOH}\) structures are therefore best regarded as conformationally related rotamers, which can interconvert by rotation about the C--O bond, while rapid pyramidal inversion at the carbon-centered radical further erases any persistent stereochemical distinction\cite{deycard1990inversion}.
Even though the transition states of these reactions cannot be isolated directly, their presence can be inferred from stereochemically distinct reaction pathways, both of which may contribute to the overall isomerization rate~\cite{danilack2021diastereomers,copan2024radical}. 
In the limiting case of degenerate transition states, their combined contribution can increase the total rate by as much as a factor of two, whereas a substantial energetic splitting can render the higher-barrier pathway kinetically irrelevant.

Oxidation chemistry of cyclic ether and peroxy-radical intermediates has been shown to involve multiple stereochemistry-dependent pathways, including competing \(\mathrm{{}^\bullet QOOH}\) isomerizations, HO$_2$ elimination channels, and ring-opening reactions leading to resonance-stabilized oxygenated radicals~\cite{doner2024stereoisomer,doner2021isomer,christianson2021reaction}. Such findings indicate that simplified stereochemically collapsed mechanisms may omit kinetically relevant channels. At the same time, large molecular databases and automated reaction workflows have greatly expanded the chemical space accessible to combustion modeling. Enumerative chemical-space efforts such as the GDB family have cataloged vast numbers of constitutional isomers and stereoisomers and have provided the foundation for various quantum-chemical datasets~\cite{fink2005virtual,fink2007virtual,blum2009970,ruddigkeit2012enumeration,buehler2026sampling,kayastha2022resolution,hoja2021qm7}. However, stereochemical variation is generally not tracked consistently along reactive coordinates, and transition states, where fleeting stereochemical motifs may become kinetically decisive, remain especially underrepresented~\cite{elliott2024role,copan2024radical,kandpal2023stereo}. Likewise, recent large-scale studies have identified many energetically accessible oxidation pathways, but stereochemical effects and diastereomeric transition states are often not treated explicitly in automated workflows~\cite{barber2021screening,elliott2024role,kandpal2023stereo}. As a result, stereochemically distinct reactive channels may be collapsed into a single effective pathway, potentially obscuring kinetically meaningful barrier differences. 
To address this gap, radical stereochemistry and diastereomeric pathways must be treated explicitly for accurate kinetic mechanism development, particularly in oxidation chemistry, where transient stereochemical distinctions can affect reactive channel counting and rate estimation~\cite{copan2024radical,elliott2024role}.

In the present work, diastereomeric transition-state pairs in \(\mathrm{ROO^\bullet \rightarrow {}^\bullet QOOH}\) isomerization are examined to determine how often such hidden stereochemical multiplicity can influence the kinetics. 
To this end, we present a new dataset, Stereochemically Expanded Autooxidation Reaction Space (SEARS), that explicitly propagates stereochemistry through radical formation, \(\mathrm{O_2}\) addition, and subsequent \(\mathrm{ROO^\bullet \rightarrow {}^\bullet QOOH}\) isomerization pathways. 
Starting from nonaromatic hydrocarbons drawn from the bigQM7$\omega$ dataset, stereochemically distinct \(\mathrm{R^\bullet}\), \(\mathrm{ROO^\bullet}\), and \({}^\bullet\mathrm{QOOH}\) intermediates are systematically generated and modeled with density functional theory (DFT). 
By quantifying how intermediate and transient stereochemical motifs reshape reaction barriers and rate constants, this work reveals a hidden kinetic dimension of hydrocarbon autooxidation chemistry and provides a stereochemically resolved reaction dataset for predictive combustion modeling and automated mechanism generation.

\section{Methods}

\subsection{Data generation}

Data generation began from the SMILES representation~\cite{o2012towards} of molecules in the bigQM7$\omega$ dataset~\cite{kayastha2022resolution}, which contains 12,880 compounds with up to seven C, N, O, and F atoms. 
This dataset was chosen as the starting point because it covers a broader portion of small-molecule chemical space than the more commonly used QM7~\cite{rupp2012fast} and QM9~\cite{ramakrishnan2014quantum} datasets. 
Specifically, bigQM7$\omega$ combines molecules drawn from both GDB11~\cite{fink2007virtual} and GDB17~\cite{ruddigkeit2012enumeration}, thereby recovering species that would be absent if either source were used alone. 
For comparison, QM7 is derived from GDB-13~\cite{blum2009970}, which itself is based on GDB-11, and contains 7,165 molecules, whereas QM9, derived from GDB17, contains only 3,993 molecules with up to seven heavy atoms. 
Molecular graphs were parsed from SMILES using RDKit~\cite{greg_landrum_2026_19250388}, and three-dimensional structures were generated in XYZ and SDF formats for subsequent calculations.

\begin{figure}[h!]
  \centering
  \includegraphics[width=1\textwidth]{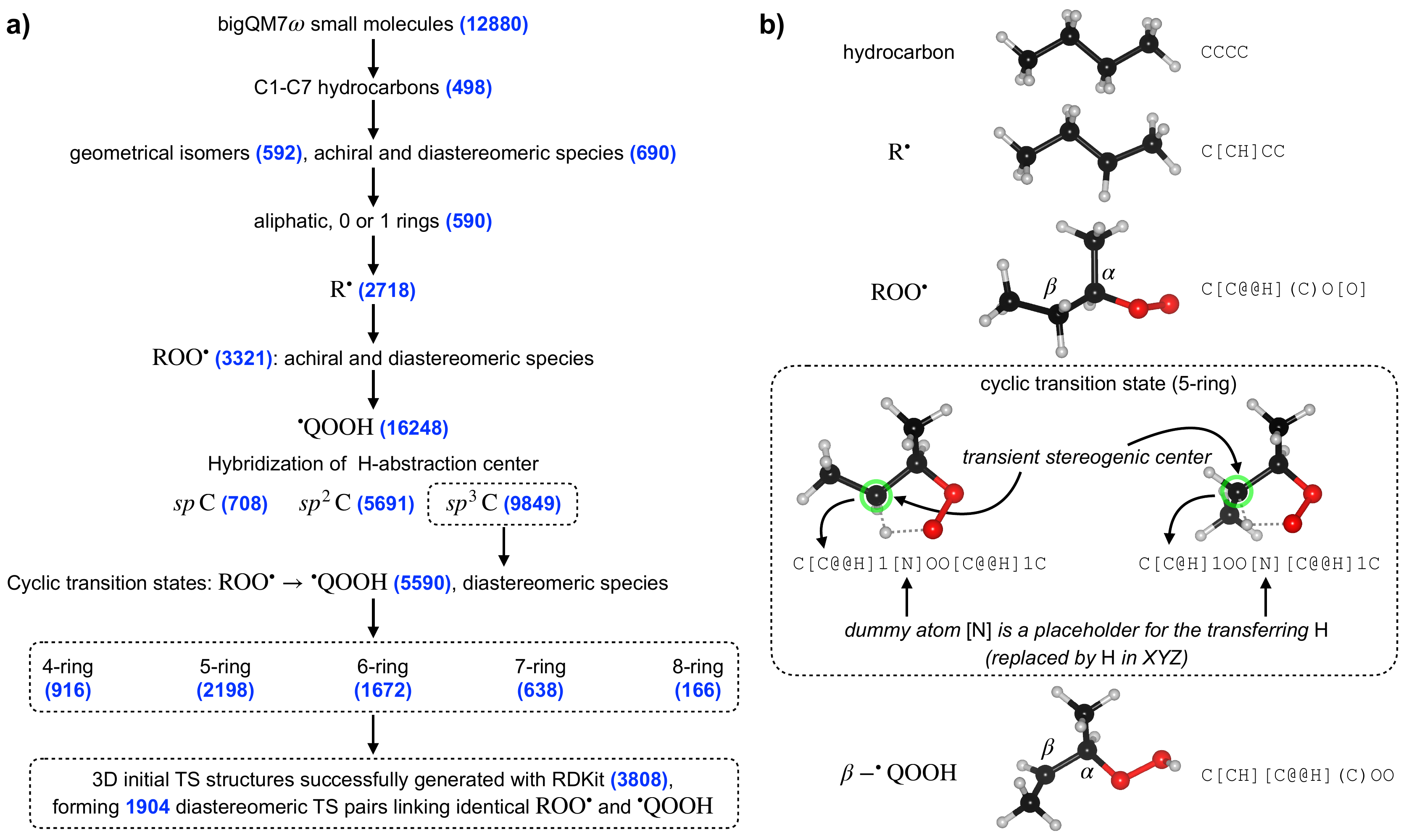}
\caption{
Workflow for generating cyclic transition states for the $\mathrm{ROO^{\bullet} \rightarrow {}^{\bullet}QOOH}$ reaction from the bigQM7$\omega$ dataset.
(a) Dataset construction and filtering procedure: From  bigQM7$\omega$, C$_1$–C$_7$ hydrocarbons were selected and filtered to retain aliphatic molecules with at most one ring. Radical and peroxy species were enumerated, yielding 3,321 $\mathrm{ROO^{\bullet}}$ radicals and 16,248 ${}^{\bullet}\mathrm{QOOH}$ species. Intramolecular H-abstraction pathways were constructed to generate 5,590 cyclic transition-state (TS) structures with ring sizes of 4-8 atoms.
(b) Representative example of the $\mathrm{ROO^{\bullet} \rightarrow {}^{\bullet}QOOH}$ reaction via a cyclic TS. A dummy atom ([N]) is used to represent the transferring hydrogen in the SMILES representation (replaced by H in the 3D structures). The H-abstraction center becomes a transient stereogenic center, giving rise to diastereomeric TS structures that connect identical $\mathrm{ROO^{\bullet}}$ reactants and ${}^{\bullet}\mathrm{QOOH}$ products. Initial 3D geometries were generated for 3,808 TS structures, corresponding to 1,904 diastereomeric TS pairs.
}
  \label{fig:datagen1}
\end{figure}

The first-level workflow for SMILES enumeration is shown in Figure~\ref{fig:datagen1}. 
From bigQM7$\omega$, 498 hydrocarbon SMILES were collected. 
Among these, benzene and toluene, the only aromatic species, were excluded. 
Further, multiply fused ring systems were also removed because such highly strained structures are not of primary interest in combustion and are expected to undergo ring opening to biradical-like structures during DFT-level geometry optimization. 
The remaining hydrocarbons were restricted to aliphatic systems that were either acyclic or contained at most one ring. These parent structures were expanded to 592 geometrical isomers by enumerating alkene \(E/Z\) stereoisomers where applicable, and then to 690 diastereomers by inversion of tetrahedral stereocenters, followed by stereochemistry-aware deduplication using canonical SMILES generated with RDKit.
This filtering and enumeration procedure yielded 590 retained diastereomers.

\begin{figure}[h!]
  \centering
  \includegraphics[width=0.7\textwidth]{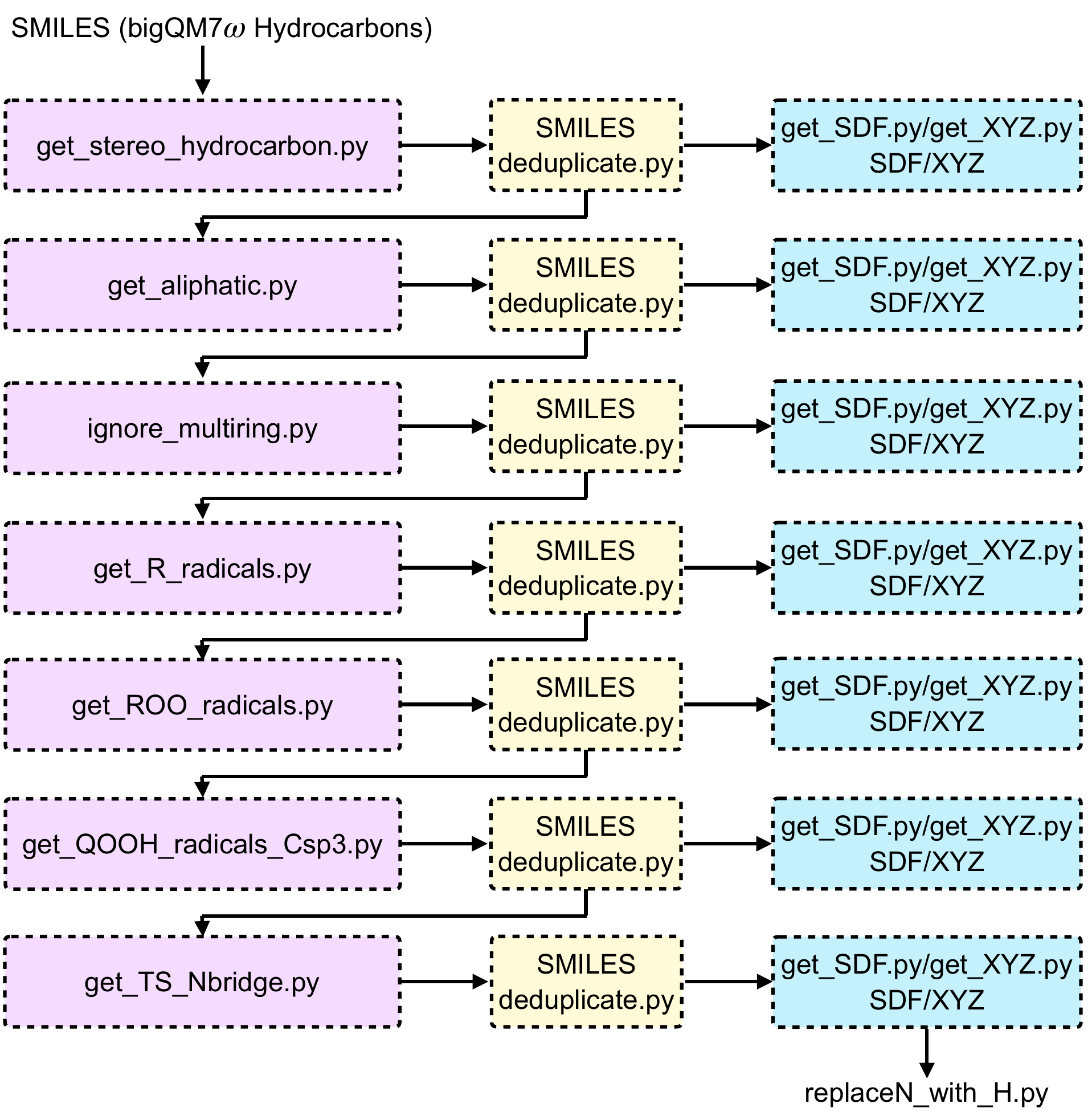}
\caption{
 Workflow used to generate stereochemically resolved SMILES of the species studied in this work, starting from the bigQM7$\omega$ hydrocarbon SMILES set. After filtering to aliphatic hydrocarbons and enumerating stereoisomeric parent structures, successive scripts were used to construct alkyl radicals, ROO radicals, and QOOH radicals, followed by generation of diastereomeric transition-state pairs for the \(\mathrm{ROO^\bullet \rightarrow {}^\bullet QOOH}\) isomerization, in which {\tt [N]} denotes the hydrogen atom abstracted by the carbon center in the QOOH product. Prior to DFT modeling, the N atom was replaced by H in the Cartesian coordinates. At each stage, duplicate SMILES were removed before conversion to SDF and XYZ formats for subsequent calculations.
}
  \label{fig:workflowSMILES}
\end{figure}

SMILES representations were generated for all possible \(\mathrm{R^\bullet}\), \(\mathrm{ROO^\bullet}\), and \({}^\bullet\mathrm{QOOH}\) radical intermediates derived from the parent hydrocarbons.  
At each stage of the SMILES-based initial structure generation workflow, duplicate SMILES were removed before conversion to SDF and XYZ formats for subsequent calculations, see Figure~\ref{fig:workflowSMILES}. Deduplication of enumerated SMILES was carried out using an enantiomer-invariant canonicalization procedure, in which each structure was represented by the lexicographically smaller of the canonical isomeric SMILES of the original molecule and its tetrahedral mirror image. 
This step eliminated redundant enantiomeric duplicates while retaining distinct diastereomers and preserving double-bond stereochemistry.

For intramolecular \(\mathrm{ROO^\bullet \rightarrow {}^\bullet QOOH}\) hydrogen-transfer reactions, SMILES representations of the 4-, 5-, 6-, 7-, and 8-membered cyclic transition-state scaffolds were generated by introducing a temporary nitrogen atom between the oxygen atom of \(\mathrm{ROO^\bullet}\) and the hydrogen-donor carbon in \({}^\bullet\mathrm{QOOH}\). 
In the SMILES representation, \texttt{[N]} was used instead of \texttt{N} so that no additional hydrogen would be introduced during conversion to three-dimensional coordinates. The nitrogen atom served only as a two-coordinate placeholder to enforce the cyclic hydrogen-transfer motif during scaffold construction.
In practice, these scaffolds were constructed by direct graph editing in RDKit, followed by stereochemical reassignment. 
This procedure enforced the cyclic topology of the hydrogen-transfer motif and uniquely specified the hydrogen atom to be abstracted. 
Nitrogen was selected as the auxiliary bridge atom after evaluating other linker atoms because, as a first-row element, it provides shorter bond lengths that better mimic the compact geometry of the hydrogen-transfer motif. 
Oxygen, although bivalent, was not used because it is already present in the \(\mathrm{ROO^\bullet}\) group and therefore would not serve as a distinct auxiliary linker that could later be removed from the XYZ structure generated from the SMILES representation. 
In contrast, heavier second-row candidates tend to form longer bonds, which distort the reaction-center geometry and make them less suitable for constructing realistic initial transition-state scaffolds. 
For example, sulfur-bridged scaffolds typically produced slightly longer C--H and O--H reaction-center distances than those found in the DFT-optimized transition-state structures for the actual hydrogen-transfer reactions, leading to a higher rate of DFT-level optimization failure.

Three-dimensional coordinates were generated from the N-bridge scaffold by RDKit distance-geometry embedding, after which the nitrogen atom was replaced by hydrogen in the Cartesian coordinates prior to any electronic-structure calculations. Transition-state optimizations were therefore carried out on the true hydrocarbon potential-energy surface, while the scaffold served only as a geometric construction device and was not retained in the quantum-chemical calculations.

From the SMILES-level enumeration, 5,590 candidate transition-state structures were generated, connecting 1,291 \(\mathrm{ROO^\bullet}\) species to 2,795 \({}^\bullet\mathrm{QOOH}\) species. 
Some of these stereochemically enumerated structures, although formally accepted as valid molecular graphs by RDKit, failed during 3D conformer generation with RDKit distance-geometry embedding. 
As a result, only 3,808 transition-state structures, corresponding to successfully generated and converged initial 3D geometries, were retained for further analysis. 
These failures occurred primarily when the transition-state ring formation generated a fused bicyclic peroxide motif, especially when the newly forming ring was fused to a pre-existing three-membered ring. 
In such cases, the connectivity and stereochemical constraints imposed incompatible geometric requirements on bond lengths and angles within the fused framework, preventing the generation of a physically reasonable 3D structure. 
Even in cases when such fused geometries could be generated, subsequent DFT optimization often cleaved the cyclopropane C--C bond, yielding a diradical rather than the intended transition-state structure.
Consequently, of the 5,590 transition-state candidates generated at the SMILES level, only 3,808 could be realized as viable 3D starting geometries for subsequent electronic-structure calculations.

All Python scripts used in the SMILES-based initial structure generation workflow are provided in the public repository described in the Data Availability section.

\subsection{Electronic structure calculations}

\begin{figure}[h!]
  \centering
  \includegraphics[width=1\textwidth]{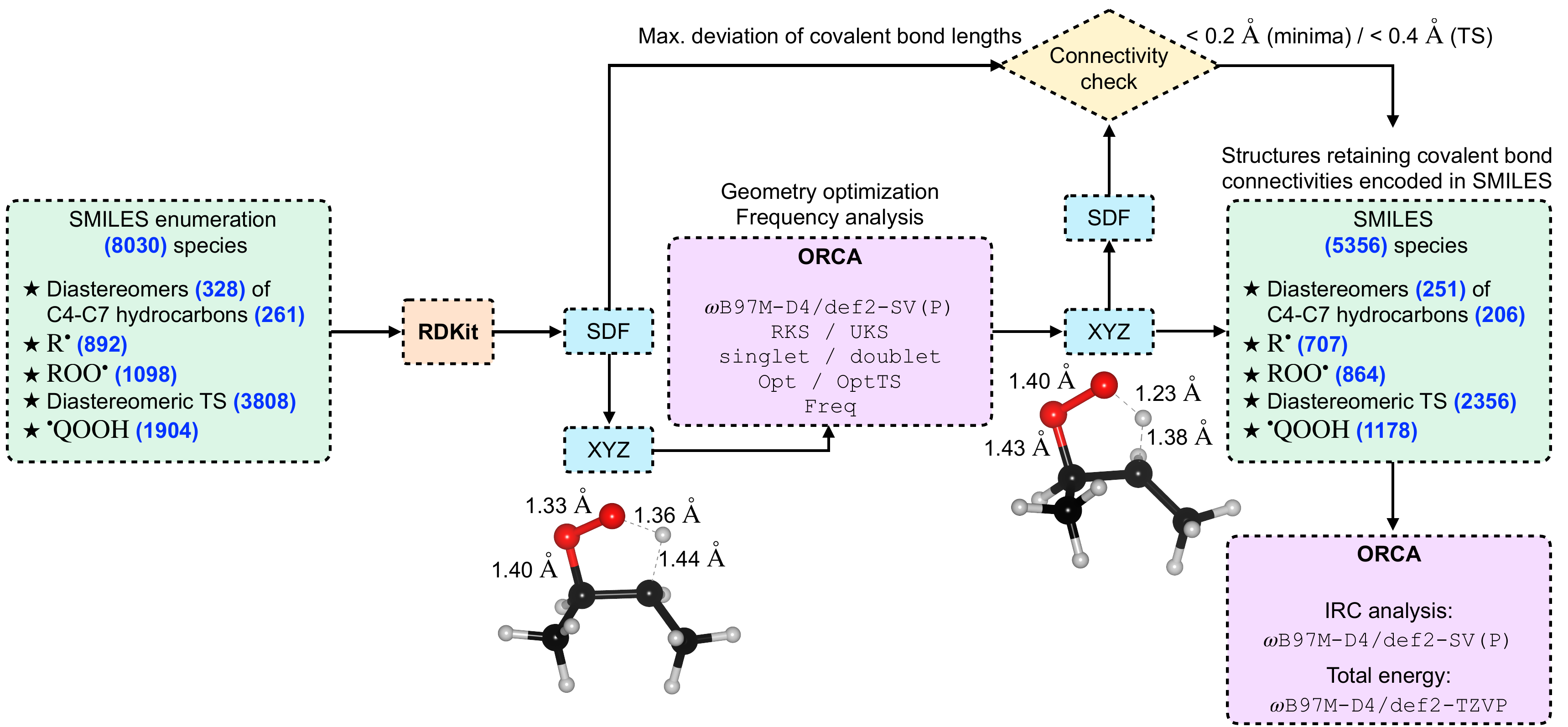}
\caption{
Workflow for the quantum chemistry data generation and characterization of diastereomeric transition states for $\mathrm{ROO^{\bullet} \rightarrow {}^{\bullet}QOOH}$ isomerization reactions. 
Starting from a chemical space of C$_4$--C$_7$ hydrocarbons, diastereomeric species were enumerated and propagated through radical ($\mathrm{R^{\bullet}}$), peroxy ($\mathrm{ROO^{\bullet}}$), and hydroperoxyalkyl (${}^{\bullet}\mathrm{QOOH}$) intermediates. 
For each \(\mathrm{ROO^{\bullet} \rightarrow {}^{\bullet}QOOH}\) channel, diastereomeric transition-state structures were optimized and frequency analyses were performed at the \(\omega\)B97M-D4/def2-SV(P) level of DFT. For a dataset of 5356 species satisfying connectivity criteria (i.e., without breaking or forming new covalent bonds), final electronic energies were computed at the \(\omega\)B97M-D4/def2-TZVP level. The workflow yields 2356 transition-state structures corresponding to 1178 \(\mathrm{ROO^{\bullet} \rightarrow {}^{\bullet}QOOH}\) reactions.
}
  \label{fig:workflowDFT}
\end{figure}

Initial geometries were generated from the SMILES-derived structures using RDKit and pre-relaxed with the UFF force field. Conformational sampling was then carried out with the GFN0-xTB~\cite{pracht2019robust} semiempirical method as implemented in xTB (version 6.7.1), using the Global Optimization Algorithm (GOAT)~\cite{de2025goat}. The {\tt GOAT-ENTROPY} keyword was employed to continue conformer searching until both the conformer energy ranking and the conformational entropy were converged, with the default threshold of \(\Delta S_{\mathrm{conf}} < 0.1\) cal/mol/K. All subsequent electronic-structure calculations were performed with ORCA (version 6.1.1)~\cite{neese2012orca,neese2025orca6}. Figure~\ref{fig:workflowDFT} summarizes the workflow used to generate the quantum-chemical dataset and to identify, optimize, and validate diastereomeric transition states for \(\mathrm{ROO^{\bullet} \rightarrow {}^{\bullet}QOOH}\) isomerization reactions.

Geometry optimizations were performed using the \(\omega\)B97M-D4 functional with the def2-SV(P) basis set. This functional was selected because the \(\omega\)B97M family has been shown to provide reliable thermochemical accuracy for radical species while remaining practical for high-throughput calculations. 
In a critical assessment against experimental enthalpies of formation for 39 free radicals in the PPE1694 dataset, \(\omega\)B97M-V and \(\omega\)B97M-D3BJ gave mean absolute deviations of 1.62 and 1.71 kcal/mol, respectively, comparable to CBS-QB3 (1.70 kcal/mol and approaching the accuracy of more expensive composite methods such as G4 (1.13 kcal/mol) and G4(MP2) (1.37 kcal/mol)~\cite{das2021critical}. 
These results indicate that the \(\omega\)B97M series provides accuracy comparable to or better than CBS-QB3 for radical thermochemistry.
We did not use \(\omega\)B97M-V~\cite{mardirossian2016omegab97m}, although it belongs to the same \(\omega\)B97M family as \(\omega\)B97M-D4. Whereas \(\omega\)B97M-V employs the nonlocal VV10 correlation functional, \(\omega\)B97M-D4 uses an empirical D4 dispersion correction. 
However, in ORCA, vibrational frequency calculations for \(\omega\)B97M-V require numerical differentiation of gradients rather than an analytic Hessian, making it less suitable for the present high-throughput workflow.

Transition states were optimized with {\tt OptTS}, and minima were optimized with {\tt Opt}. 
Transition-state structures were confirmed by the presence of a single imaginary vibrational frequency, whereas minima were verified to have no imaginary frequencies. 
To improve optimization robustness, Hessians were recomputed every five geometry cycles to converge toward minima in {\tt Opt} calculations and first-order saddle points in {\tt OptTS} calculations.
A maximum of 100 geometry iterations was allowed for all optimizations. 
All calculations employed the {\tt RIJCOSX} approximation with the {\tt DEF2/J} auxiliary basis set and the {\tt TightSCF} convergence criterion. Open-shell radical intermediates and transition states were treated with unrestricted Kohn--Sham (UKS) calculations and a doublet spin multiplicity, whereas closed-shell hydrocarbons were treated as singlets.

Intrinsic reaction coordinate (IRC) calculations were also carried out at the \(\omega\)B97M-D4/def2-SV(P) level to verify that each transition state connected the intended reactant and product channels. IRC calculations were typically performed with 100 steps, using {\tt irc maxiter 100} as the maximum number of iterations per step. 
Although this setting was not always sufficient to relax every trajectory fully to its endpoint, the purpose of the IRC calculations here was to confirm the reaction channel and exclude transition states leading to undesired products rather than the target \({\rm ROO}^\bullet\) or \({}^\bullet{\rm QOOH}\) species. 
Because of the large number of systems considered, IRC endpoints were not further optimized individually; instead, connectivity checks were used to confirm that the IRC paths evolved toward the expected products.

Final electronic energies were obtained from \(\omega\)B97M-D4/def2-TZVP single-point calculations on \(\omega\)B97M-D4/def2-SV(P)-optimized geometries. Vibrational analysis, including zero-point and thermal corrections, was carried out at the \(\omega\)B97M-D4/def2-SV(P) level.

\section{Results and Discussion}

\subsection{IRC validation of \(\mathrm{ROO^{\bullet} \rightarrow {}^{\bullet}QOOH}\) transition states}

\begin{figure}[h!]
  \centering
  \includegraphics[width=1.0\linewidth]{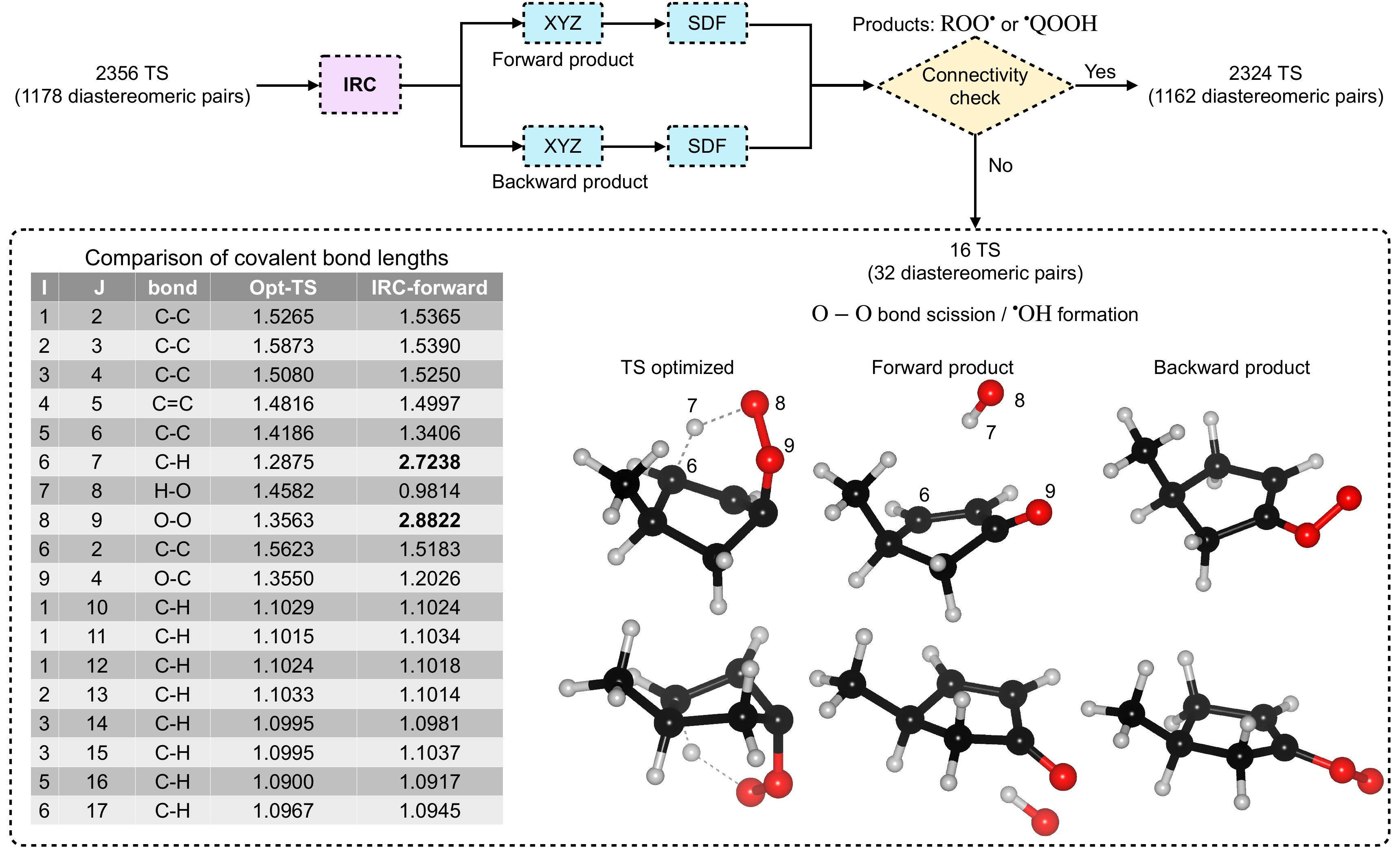}
\caption{Workflow used to diagnose connectivity mapping in 2,356 IRC calculations for 1,178 diastereomeric pairs of transition states associated with $\mathrm{ROO^{\bullet} \rightarrow {}^{\bullet}QOOH}$ reactions at the $\omega$B97M-D4/def2-SV(P) level of theory. The SDF file generated from each optimized TS structure, using the cyclic TS initial connectivity, was compared with the SDF files generated from the final forward and backward IRC structures (100 IRC steps in total) to determine whether the IRC connects the intended $\mathrm{ROO^{\bullet}}$ and ${}^{\bullet}\mathrm{QOOH}$ minima. Most IRC calculations (2,324 TS; 1,162 diastereomeric pairs) led to the expected $\mathrm{ROO^{\bullet}}$ and ${}^{\bullet}\mathrm{QOOH}$ products, whereas a small subset led to an alternate reaction pathway involving O--O bond scission and ${}^{\bullet}\mathrm{OH}$ formation. Representative failed cases are shown together with a comparison of covalent bond lengths in the optimized TS and IRC endpoint structures.}
\label{fig:ircworkflow}
\end{figure}

An important step in dataset construction was verification that the optimized first-order saddle points actually connect the intended \(\mathrm{ROO^{\bullet}}\) and \({}^{\bullet}\mathrm{QOOH}\) minima. 
For this purpose, IRC calculations were carried out from each optimized transition state, and the resulting forward and backward endpoint structures were compared against the target reactant and product connectivities, as illustrated in Figure~\ref{fig:ircworkflow}. 
Here, the IRC calculations were used primarily as a diagnostic tool to identify transition states that drifted into unintended reaction channels rather than to obtain fully converged endpoint structures, as only a limited number of IRC steps were performed.

Of the 3,808 RDKit-level initial transition-state geometries generated for \(\mathrm{ROO^{\bullet} \rightarrow {}^{\bullet}QOOH}\) isomerization, 2,356 DFT-optimized structures, corresponding to 1,178 diastereomeric pairs, retained the intended covalent bonding connectivity and were taken forward for IRC validation. Among these, 2,324 structures, corresponding to 1,162 diastereomeric pairs, were confirmed by IRC analysis to proceed along the intended forward and reverse paths toward the \(\mathrm{ROO^{\bullet}}\) and \({}^{\bullet}\mathrm{QOOH}\) minima. The remaining 32 structures, corresponding to 16 diastereomeric pairs, instead evolved toward an alternative channel involving O--O bond scission and \({}^{\bullet}\mathrm{OH}\) formation. Thus, although these structures satisfied the vibrational criterion for a first-order saddle point, they were excluded from the final \(\mathrm{ROO^{\bullet} \rightarrow {}^{\bullet}QOOH}\) dataset because they did not connect the target reactant and product wells.

An important common feature of all failed cases is that the peroxyl group is attached to an \(sp^2\) carbon center. This suggests that such systems are particularly susceptible to geometric and electronic distortion along the hydrogen-transfer coordinate. In these structures, orienting the OO group toward the hydrogen-abstraction site introduces additional strain at the \(sp^2\) carbon, which appears to promote access to competing O--O scission channels rather than the intended intramolecular H-transfer pathway. These structures should therefore not be regarded as failed \(\mathrm{ROO^{\bullet} \rightarrow {}^{\bullet}QOOH}\) transition states, but rather as valid saddle points leading to a different region of the potential-energy surface.

Structure names, SMILES representations, electronic energies, and IRC-related descriptors for the 2,356 DFT-optimized transition states and the 2,324 IRC-validated \(\mathrm{ROO^{\bullet} \rightarrow {}^{\bullet}QOOH}\) transition states are provided in the public repository (see Data Availability) as separate CSV files. The corresponding Cartesian coordinates are provided separately in folders named according to the structure identifiers reported in the CSV entries.

\subsection{Energetic trends in diastereomeric transition state pairs}

\begin{figure}[h!]
  \centering
  \includegraphics[width=0.7\textwidth]{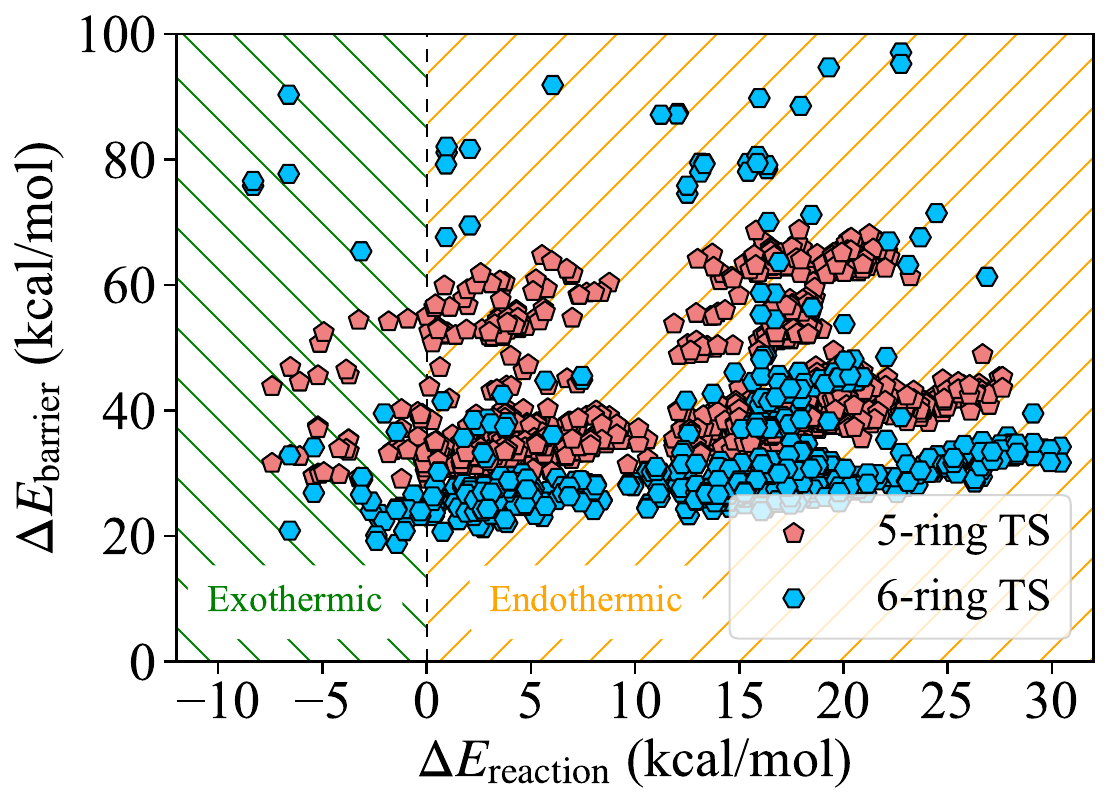}
\caption{
Joint distribution of reaction energies and forward barrier heights for 1,162 $\mathrm{ROO^{\bullet} \rightarrow {}^{\bullet}QOOH}$ isomerization reactions. Each reaction is represented by two points corresponding to a pair of diastereomeric transition-state structures of the same ring size.
}
  \label{fig:eyringplot}
\end{figure}

\begin{figure}[h!]
  \centering
  \includegraphics[width=0.7\textwidth]{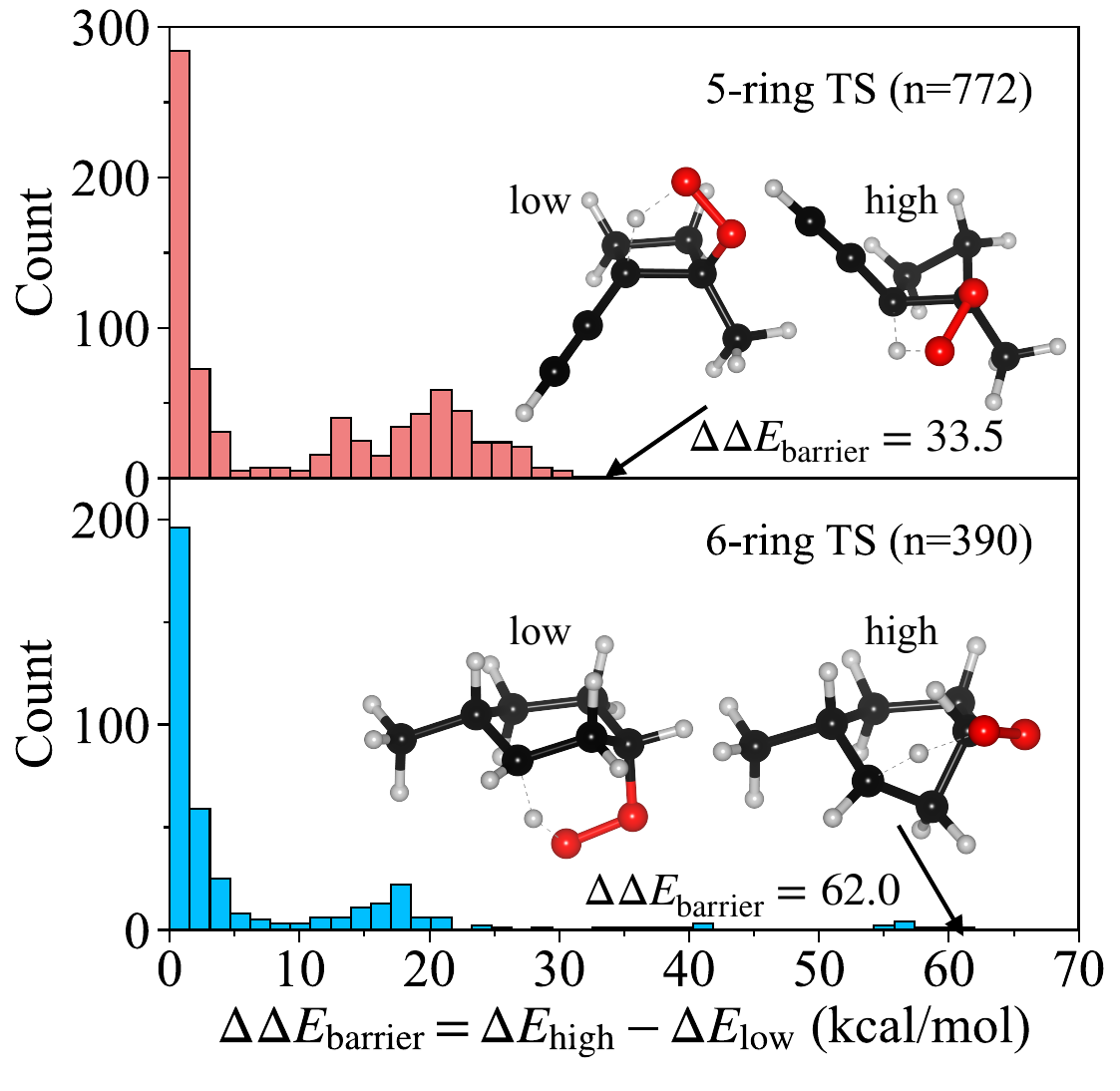}
\caption{
Histogram of difference in electronic energies between 
diastereomeric transition states
($\Delta\Delta E_{\rm barrier} = \Delta E_{\rm high} - \Delta E_{\rm low}$) 
for 1162
$\mathrm{ROO^{\bullet} \rightarrow {}^{\bullet}QOOH}$ reactions derived from
856 unique $\mathrm{ROO^{\bullet}}$ species. 
Representative
transition-state structures for maximum $\Delta\Delta E_{\rm barrier}$
for each category are shown.
}
  \label{fig:histogram}
\end{figure}

\begin{figure}[h!]
  \centering
  \includegraphics[width=0.7\textwidth]{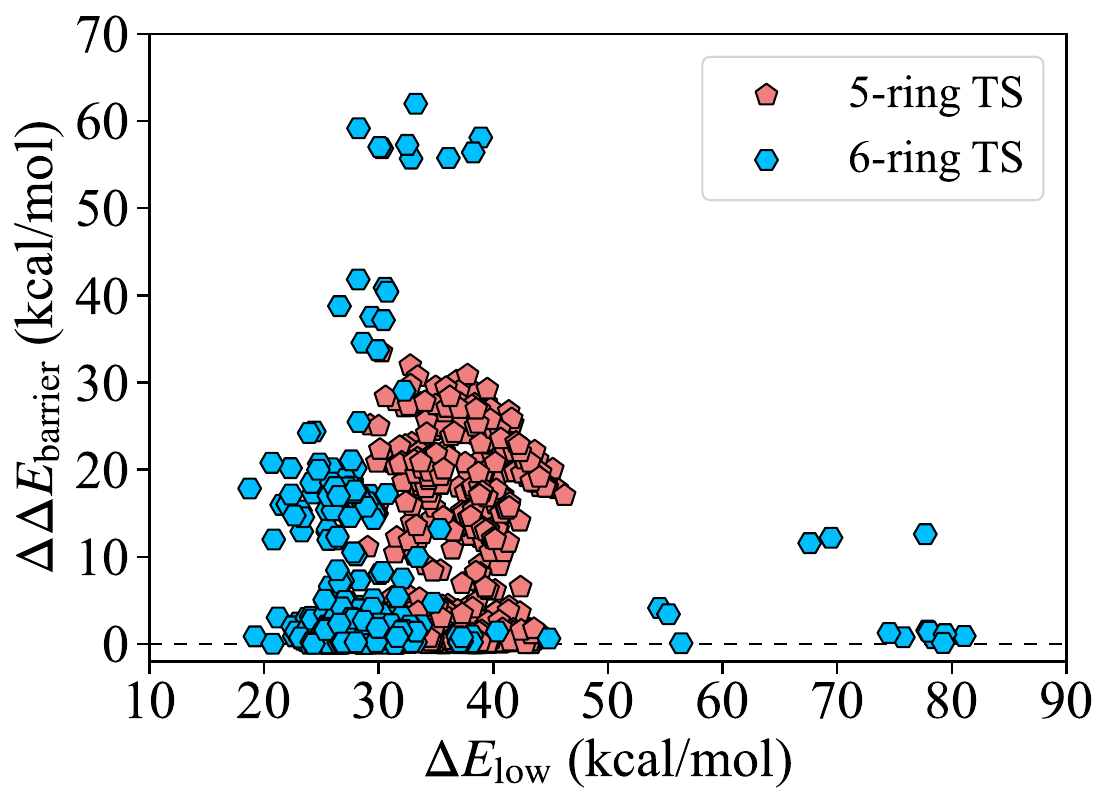}
\caption{
Energy difference between diastereomeric transition states, \(\Delta\Delta E_{\rm barrier} = \Delta E_{\rm high} - \Delta E_{\rm low}\), plotted against the lower barrier height, \(\Delta E_{\rm low}\), for 1162 \(\mathrm{ROO^{\bullet} \rightarrow {}^{\bullet}QOOH}\) isomerization reactions.
}
  \label{fig:scatterddEdlow}
\end{figure}

\begin{figure}[h!]
  \centering
  \includegraphics[width=1.0\linewidth]
{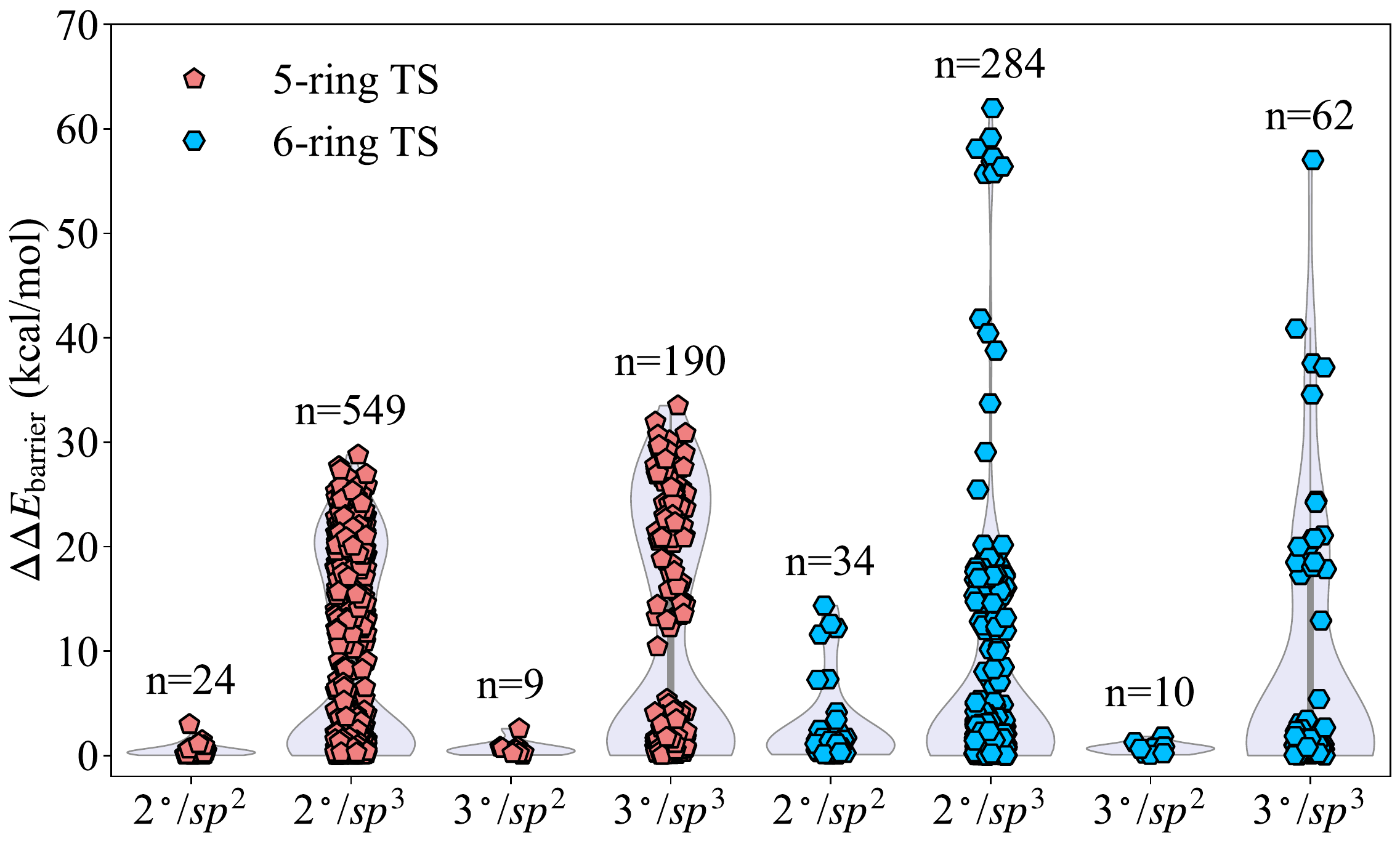}
\caption{
Violin-plot distribution of diastereomeric barrier differences for
$\mathrm{ROO^{\bullet} \rightarrow {}^{\bullet}QOOH}$  isomerization, $\Delta\Delta E_{\rm barrier}$, grouped by transition-state ring size, degree of substitution at the H-abstraction $sp^3$ carbon site ($2^\circ$, $3^\circ$), and hybridization of the carbon bonded to the peroxyl group 
($sp^2$, $sp^3$). Data points corresponding to  
5- and 6-membered transition states are shown as
pentagons and hexagons, respectively. 
}
\label{fig:violinplot}
\end{figure}

The energetic landscape of the validated diastereomeric transition-state dataset is summarized in Figures~\ref{fig:eyringplot}--\ref{fig:violinplot}. Each \(\mathrm{ROO^{\bullet} \rightarrow {}^{\bullet}QOOH}\) reaction is associated with a pair of diastereomeric transition states that connect the same reactant and product wells but differ in the stereochemical arrangement along the hydrogen-transfer coordinate. The extent to which these paired transition states differ in energy provides a direct measure of the stereochemical sensitivity of the barrier for intramolecular H-shift.

Figure~\ref{fig:eyringplot} shows the joint distribution of reaction energies and forward barrier heights for the full set of validated reactions. The two members of each diastereomeric pair share the same reaction energy and therefore align at the same position along the horizontal axis, reflecting their common reactant and product wells and hence the same overall thermochemical driving force. At the same time, the separation of paired points along the vertical axis shows that, even for a fixed reaction energy, the forward barrier can depend appreciably on the relative three-dimensional arrangement of the transferring hydrogen, peroxyl group, and surrounding substituents. Thus, stereochemistry does not merely generate duplicate representations of the same reaction coordinate, but can give rise to  distinct transition-state structures with measurably different barrier heights.

The distribution of these differences is shown explicitly in Figure~\ref{fig:histogram}, which reports \(\Delta\Delta E_{\rm barrier} = \Delta E_{\rm high} - \Delta E_{\rm low}\) for each diastereomeric pair. The histogram is dominated by pairs with relatively small \(\Delta\Delta E_{\rm barrier}\), indicating that near-degenerate diastereomeric pathways are common in this dataset. 
The predominance of near-degenerate diastereomeric pathways is also evident from the cumulative distribution of \(\Delta\Delta E_{\rm barrier}\). Of the 1,162 validated diastereomeric transition-state pairs, 372 (32.0\%) have \(\Delta\Delta E_{\rm barrier} \leq 1\) kcal/mol, 541 (46.6\%) lie within 2 kcal/mol, and 604 (52.0\%) lie within 3 kcal/mol. The cumulative fractions increase to 649 (55.9\%) and 673 (57.9\%) within 4 and 5 kcal/mol, respectively. These statistics show that small barrier splittings are common in the dataset, even though a substantial tail of larger \(\Delta\Delta E_{\rm barrier}\) values remains and gives rise to strongly differentiated kinetic behavior.
In these cases, one member of the pair accesses a comparatively favorable hydrogen-transfer geometry, whereas the other is destabilized by increased strain, less favorable ring puckering, or greater geometric distortion at the incipient transition-state ring. The representative structures shown for the largest values of \(\Delta\Delta E_{\rm barrier}\) illustrate that these large splittings are typically associated with visibly different local strain patterns. 

The relationship between the magnitude of the diastereomeric splitting and the absolute barrier height is further clarified in Figure~\ref{fig:scatterddEdlow}. 
Although both members of a given diastereomeric pair share the same reaction energy, the distribution of points across the horizontal axis shows that different classes of \(\mathrm{ROO^{\bullet} \rightarrow {}^{\bullet}QOOH}\) reactions occupy distinct thermochemical regimes. 
In particular, reactions proceeding through 6-membered transition states are concentrated mainly at lower reaction energies, typically around 20--30 kcal/mol, whereas 5-membered transition states are more commonly found at higher reaction energies, around 30--40 kcal/mol. 
A smaller number of 6-membered cases also extend to reaction energies above 50 kcal/mol, while retaining somewhat similar barrier heights. 
This distribution is consistent with prior work showing that the stability and reactivity of ${\rm \bullet QOOH}$-related species are strongly influenced by stereoelectronic effects and local geometric constraint~\cite{kandpal2023stereo}.

A clearer view of structural trends is provided by the class-resolved distributions in Figure~\ref{fig:violinplot}, where \(\Delta\Delta E_{\rm barrier}\) is grouped according to transition-state ring size, the degree of substitution at the H-abstraction carbon, and the hybridization of the carbon bearing the peroxyl group. 
In general, systems in which the peroxyl group is attached to an \(sp^2\) carbon tend to show smaller \(\Delta\Delta E_{\rm barrier}\) values than systems in which the peroxyl group is attached to an \(sp^3\) carbon. This suggests that the local geometry near an \(sp^2\)-substituted peroxyl center may limit the extent to which the two diastereomeric arrangements can diverge energetically. However, this subset is smaller, and the observed spread, especially for cases involving a secondary H-abstraction site, indicates that broader variation may emerge as additional examples are sampled. 
The largest \(\Delta\Delta E_{\rm barrier}\) values are observed for systems in which the peroxyl group is attached to an \(sp^3\) carbon and the hydrogen is abstracted from a secondary carbon center. These cases appear to be especially sensitive to stereochemical rearrangement because inversion at the H-abstraction site can force one diastereomeric transition state into a much less favorable spatial arrangement than the other.

This effect is particularly pronounced when the peroxyl-bearing carbon is embedded in or adjacent to a ring. In such cases, the cyclic transition-state motif must be achieved without the same degree of conformational flexibility available in acyclic systems. 
As a result, one diastereomer may adopt a relatively low-strain arrangement, whereas the other must accommodate substantial eclipsing or unfavorable across-ring nonbonded interactions, which destabilize the transition-state structure and raise the energy barrier.
The largest observed splittings therefore arise not simply from the presence of stereocenters, but from the way stereochemistry couples to ring constraint and local crowding in the transition-state geometry.

\subsection{Implications for combustion kinetics and automated mechanism generation}

For a reaction proceeding via a transition state, transition-state theory gives the unimolecular rate constant as
\begin{equation}
k=\kappa \frac{k_{\mathrm B}T}{h}\exp\!\left(-\frac{\Delta G^\ddag}{RT}\right),
\end{equation}
where \(\Delta G^\ddag\) is the activation free energy and \(\kappa\) is the transmission coefficient, i.e., the probability that an activated complex proceeds to products rather than recrossing to reactants~\cite{eyring1935activated}. When multiple symmetry-equivalent reactive arrangements exist, the rate constant must additionally include the appropriate reaction-path degeneracy, or equivalently the corresponding symmetry-number correction\cite{schlag1963symmetry}. Thus, if one transition-state geometry represents one of \(g\) equivalent reactive arrangements, the observable rate constant becomes
\begin{equation}
k=g\,\kappa \frac{k_{\mathrm B}T}{h}\exp\!\left(-\frac{\Delta G^\ddag}{RT}\right).
\end{equation}
For ordinary cases, the reaction-path symmetry number is given by the ratio of the total symmetry number of the reactant to that of the transition state\cite{fernandez2007symmetry}. In practice, however, the assignment of symmetry corrections or statistical factors is not always straightforward\cite{pollak1978symmetry}, particularly for symmetric reactions, chiral species, multiple conformers, and internal rotations\cite{fernandez2007symmetry}.

Diastereomeric transition states are fundamentally different from symmetry-equivalent pathways. If two diastereomeric transition states connect the same reactant and product wells, the total rate constant is obtained by summing the two parallel contributions,
\begin{equation}
k_{\mathrm{tot}}=k_{\mathrm{low}}+k_{\mathrm{high}}.
\end{equation}
It then follows that
\begin{equation}
\frac{k_{\mathrm{tot}}}{k_{\mathrm{low}}}
=
1+\exp\!\left(-\frac{\Delta\Delta G^\ddag}{RT}\right),
\end{equation}
where
\begin{equation}
\Delta\Delta G^\ddag=\Delta G^\ddag_{\mathrm{high}}-\Delta G^\ddag_{\mathrm{low}}\ge 0.
\end{equation}
This expression is a direct consequence of transition-state theory for parallel pathways and reflects the more general principle that the overall rate is obtained by summing the contributions of all transition states that connect the same reactant and product wells\cite{burke2017ephemeral}.

\noindent
\begin{figure}[h!]
  \centering
  \includegraphics[width=0.7\textwidth]{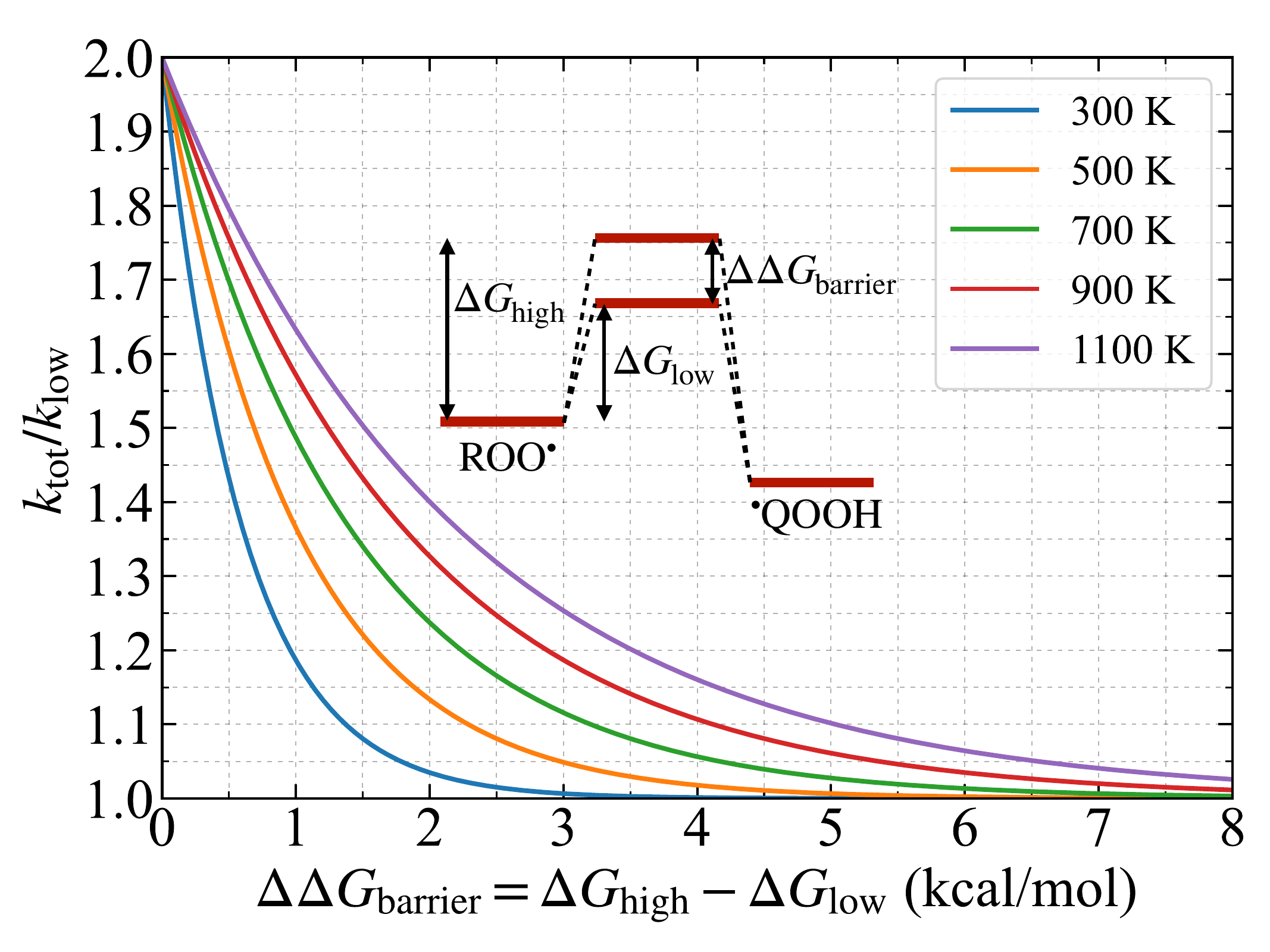}
\caption{
Ratio of the total rate constant for two parallel pathways, $k_{\mathrm{tot}} = k_{\mathrm{low}} + k_{\mathrm{high}}$, to the rate constant of the lower-barrier pathway, $k_{\mathrm{low}}$, as a function of the barrier free-energy difference, $\Delta\Delta G_{\mathrm{barrier}} = \Delta G_{\mathrm{high}} - \Delta G_{\mathrm{low}}$, at temperatures from 300 to 1100 K. The inset shows a schematic free-energy profile defining $\Delta\Delta G_{\mathrm{barrier}}$ for two competing transition states connecting the same reactant and product wells.
}
  \label{fig:rateplot}
\end{figure}

Figure~\ref{fig:rateplot} illustrates the kinetic consequence of this relation. When the two diastereomeric transition states are nearly degenerate, both pathways contribute substantially and the total rate approaches twice that of the lower-barrier pathway alone. As the free-energy splitting increases, the higher-energy diastereomer contributes progressively less, and the total rate rapidly approaches \(k_{\mathrm{low}}\). The figure therefore provides a simple quantitative criterion for judging when a second diastereomeric transition state is kinetically relevant.

The relations above are stated in terms of activation free energies because the kinetic consequences of competing diastereomeric pathways are formally governed by transition-state theory. In the representative examples discussed below, however, the comparison is made using electronic energies rather than free energies, in order to highlight the underlying barrier splitting and conformational structure of the computed stationary points. These examples are therefore intended to illustrate the qualitative kinetic consequences of diastereomeric splitting, rather than to provide fully quantitative rate predictions.

\subsection{Representative cases and kinetic consequences}

\begin{figure}[h!]
  \centering
  \includegraphics[width=1.0\linewidth]{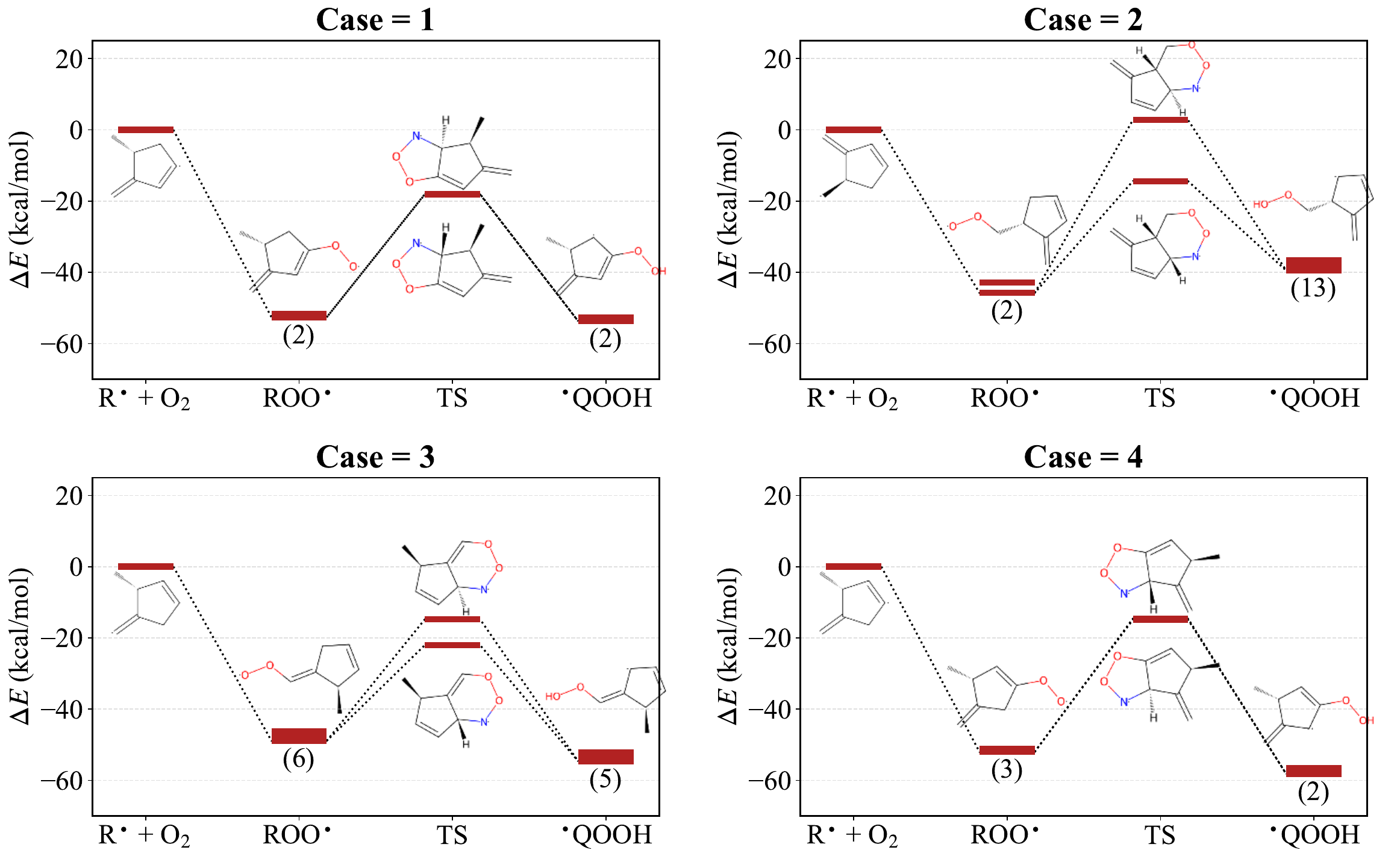}
\caption{
Energy profiles for selected \(\mathrm{R^\bullet + O_2 \rightarrow ROO^\bullet \rightarrow {}^\bullet QOOH}\) reactions, showing the relative energies of \(\mathrm{R^\bullet + O_2}\), all located \(\mathrm{ROO^\bullet}\) conformers, the two diastereomeric transition-state structures, and all located \({}^\bullet\mathrm{QOOH}\) conformers. Numbers in parentheses indicate the number of conformers identified for the corresponding \(\mathrm{ROO^\bullet}\) and \({}^\bullet\mathrm{QOOH}\) wells. In the transition-state drawings, N is used in place of the abstracted H atom to highlight the H-transfer site.
}
\label{fig:conf2x2}
\end{figure}

To illustrate the kinetic consequences of diastereomeric barrier splitting most clearly, representative exothermic \(\mathrm{ROO^\bullet \rightarrow {}^\bullet QOOH}\) reactions were examined in detail. 
Among the 1162 isomerization reactions examined in this work, only 41 are exothermic with respect to the corresponding \(\mathrm{ROO^\bullet}\) and \({}^\bullet\mathrm{QOOH}\) minima. 
Figure~\ref{fig:conf2x2} presents energy-profile diagrams for four representative examples, while the corresponding profiles for all 41 reactions are presented in 
Figure~S1 of 
the Supporting Information. 
For each case, the diagrams include the energy of \(\mathrm{R^\bullet + O_2}\), all located \(\mathrm{ROO^\bullet}\) conformers, the two diastereomeric transition-state structures, and all located \({}^\bullet\mathrm{QOOH}\) conformers. 
In most cases, only a small number of low-lying conformers were identified, although one \({}^\bullet\mathrm{QOOH}\) example exhibited 13 conformers within a narrow energy window, indicating a relatively flat product-side conformational landscape of this system.

The four examples were selected to illustrate how the energy splitting between diastereomeric transition structures can vary even within this small exothermic subset. 
Cases~1 and~2 originate from the same hydrocarbon framework but differ in the site of \(\mathrm{O_2}\) addition that forms the corresponding \(\mathrm{ROO^\bullet}\) intermediate. 
Likewise, Cases~3 and~4 are derived from the same hydrocarbon scaffold but differ in the radical site at which \(\mathrm{O_2}\) addition occurs. 
These comparisons show that relatively subtle changes in \(\mathrm{ROO^\bullet}\) connectivity can significantly affect both the magnitude of the forward barrier and the energetic separation between the two diastereomeric transition structures.

Cases~1 and~4 proceed through five-membered-ring transition states and exhibit only very small energy differences within the diastereomeric TS pair, with forward barriers of about 34 kcal/mol in Case~1 and about 36 kcal/mol in Case~4. 
In contrast, Cases~2 and~3 proceed through six-membered-ring transition states and display substantially larger splitting within the diastereomeric TS pair. 
For Case~2, the lower- and higher-energy barriers are approximately 28 and 46 kcal/mol, respectively, whereas for Case~3 the corresponding barriers are about 24 and 31 kcal/mol. 
These examples therefore show that even within exothermic \(\mathrm{ROO^\bullet \rightarrow {}^\bullet QOOH}\) reactions, the two diastereomeric pathways can differ markedly in accessibility.

The lower barriers observed for Cases~2 and~3 relative to their high barrier counterparts, Cases~1 and~4, are broadly consistent with previous combustion studies showing that six-membered H-shift transition states are often more favorable than five-membered analogues~\cite{villano2012high}. 
More generally, the present ring-size and local-structure trends are consistent with the broader rate-rule perspective that intramolecular peroxy-radical H-shifts depend systematically on the topology of the migrating hydrogen and the local bonding environment around the peroxyl center~\cite{villano2012high}. 
At the same time, the present examples should not be interpreted as implying a simple one-to-one relationship between ring size and either barrier height or diastereomeric splitting. 
As shown by the broader dataset, both five- and six-membered transition states can exhibit either small or large diastereomeric energy differences. In addition, stereochemical and topological effects can modify the expected ring-size trend; for example, five-membered transition states with spiro-type attachment to another ring can in some cases be lower in energy than six-membered fused-ring transition states~\cite{kandpal2023stereo}. 
The representative profiles shown here therefore emphasize that the extent of diastereomeric splitting is system-dependent and reflects a combination of local substitution pattern, H-abstraction geometry, and conformational preferences of the reactant and product wells.

The barrier magnitudes obtained here are also consistent with the broad range reported previously for combustion-relevant \(\mathrm{ROO^\bullet \rightarrow {}^\bullet QOOH}\) isomerizations. 
For example, Danilack \emph{et al.}\ reported QOOH-forming barriers of 19.04, 24.79, and 36.91 kcal/mol for 1-ethoxyethyl-derived peroxy radicals~\cite{danilack2021diastereomers}, while lower H-shift barriers, including values below 20 kcal/mol, have been reported for some acyclic alkylperoxy systems such as \(n\)-pentyl oxidation chemistry~\cite{asatryan2010chain}. 
The present set therefore spans both moderately accessible and relatively inaccessible exothermic pathways, with the 46 kcal/mol barrier in Case~2 lying toward the upper end of the range relevant to \(\mathrm{ROO^\bullet \rightarrow {}^\bullet QOOH}\) isomerization.

For a pair of diastereomeric transition states connecting the same \(\mathrm{ROO^\bullet}\) well to the same \({}^\bullet\mathrm{QOOH}\) well, the two pathways contribute in parallel in terms of rate constants rather than through any direct combination of barrier heights. 
In near-degenerate cases, this may effectively reduce to a degeneracy correction, whereas for unequal barriers the lower- and higher-energy pathways contribute unequally. 
If both barriers are high, the overall isomerization may be uncompetitive with other available reaction channels. 
In the limiting case where one barrier is much higher than the other, the higher-barrier pathway becomes kinetically negligible.

Since the rate constants depend exponentially on the barrier height, even modest diastereomeric splittings can strongly bias the kinetics toward the lower-energy pathway. 
The four representative cases considered here illustrate both limiting behaviors. 
In Cases~1 and~4, the two diastereomeric barriers are nearly degenerate, so both pathways are expected to contribute comparably to the total isomerization rate. 
In comparison, Case~2 exhibits a barrier splitting of about 18 kcal/mol, implying that the higher-energy pathway should be kinetically negligible relative to the lower-energy one. 
Case~3 shows an intermediate situation, where although the splitting is smaller (about 7 kcal/mol), the lower-energy TS is still expected to dominate the overall rate. 
Thus, the kinetic importance of both members of a diastereomeric TS pair is greatest when their barriers are both low enough for \(\mathrm{ROO^\bullet \rightarrow {}^\bullet QOOH}\) isomerization to compete with other reaction channels and similar enough that both pathways are favored.

The class-resolved analysis shows that the magnitude of the diastereomeric splitting depends systematically on local structure, including the transition-state ring size, the substitution at the H-abstraction center, and the hybridization of the peroxyl-bearing carbon. In particular, large splittings arise when stereochemistry couples strongly to ring constraint and local steric crowding, whereas near-degenerate pairs are more common in less differentiated environments. This suggests that stereochemistry-aware mechanism generation may be especially important for cyclic or conformationally restricted systems, where one diastereomeric pathway can be selectively destabilized, and for cases where multiple nearly degenerate pathways increase the total rate.

Overall, the results highlight a limitation of automated reaction-discovery workflows that propagate constitutional connectivity but do not consistently retain stereochemical information at the level of transition states. Because \(\mathrm{O_2}\) addition, H-transfer, and ring-forming reactions can generate new stereocenters or transient stereogenic environments, automated workflows that collapse such distinctions may undercount both the number of accessible channels and their combined kinetic contribution. The present SEARS framework therefore provides a route toward stereochemically aware reaction-network construction, in which reactants, products, and transition states are represented at a level sufficient to distinguish kinetically distinct diastereomeric pathways. The present results support recent arguments that stereochemistry should be represented explicitly in automated kinetic mechanism development, because constitutionally identical radicals can access multiple diastereomeric pathways with distinct kinetic weights\cite{copan2024radical,elliott2024role,moller2019importance}.

\section{Conclusions}

In this work, we introduced a Stereochemically Expanded Autooxidation Reaction Space (SEARS) for low-temperature hydrocarbon oxidation and used it to examine diastereomeric transition states in \(\mathrm{ROO^\bullet \rightarrow {}^\bullet QOOH}\) isomerization. By propagating stereochemistry from hydrocarbons to \(\mathrm{R^\bullet}\), \(\mathrm{ROO^\bullet}\), and \({}^\bullet\mathrm{QOOH}\) species, and by constructing stereochemically resolved cyclic transition-state guesses, we showed that a constitutional-only description can miss reactive channels that are stereochemically distinct and kinetically important.

Density functional theory calculations together with IRC-based connectivity checks identified a large set of validated diastereomeric transition-state pairs connecting the same \(\mathrm{ROO^\bullet}\) and \({}^\bullet\mathrm{QOOH}\) wells. The results reveal two chemically important regimes. In many cases, the two diastereomeric transition states are close in energy and may both contribute appreciably to the overall rate. In other cases, substantial barrier splitting is observed, such that one pathway is expected to dominate kinetically while the other becomes effectively negligible. Stereochemical multiplicity is therefore not merely a labeling issue, but a meaningful source of barrier diversity and, in near-degenerate cases, rate enhancement.

The class-resolved analysis further shows that the magnitude of diastereomeric splitting depends systematically on local structure, including transition-state ring size, substitution at the H-abstraction center, and the hybridization of the peroxyl-bearing carbon. In particular, large splittings arise when stereochemistry couples strongly to ring constraint and local steric crowding, whereas near-degenerate pairs are more common in less differentiated environments. These trends provide a chemical-space perspective on where explicit stereochemical resolution is most likely to matter in combustion kinetics.

Overall, this study highlights the importance of treating transition-state stereochemistry explicitly in automated reaction discovery and mechanism generation. Datasets that represent only a single optimized structure per constitutional formula are likely to underrepresent the kinetic complexity of hydrocarbon autooxidation. In this sense, the present SEARS framework and curated quantum-chemical dataset provide a foundation for stereochemically aware rate estimation, mechanism development, and future data-driven models of combustion reactivity. Stereochemistry-aware graph formalisms such as StereoMolGraph may be especially useful in this context because they can encode reactant, product, and transition-state stereochemistry within a unified reaction representation~\cite{papusha2026stereomolgraph}.

Although further refinement using higher-level electronic-structure methods, more exhaustive conformational sampling, extended reaction-path validation, and fully free-energy-based kinetic analysis will be valuable for reaction-specific quantitative predictions, the present results already demonstrate that diastereomeric pathway splitting is sufficiently common in this chemical space to warrant explicit consideration in future combustion workflows. Beyond its mechanistic significance, the dataset also provides a useful benchmark for future machine-learning studies of combustion kinetics, particularly for models aimed at learning not only average barrier trends but also the stereochemically induced splitting and near-degeneracy patterns that govern competing pathways~\cite{flora2024prediction,johnson2024machine,zhang2025enhancing,alfonso2024repurposing}.

\begin{acknowledgement}
  This work was supported by the 
Department of Atomic Energy, Government
of India, under Project Identification No.~RTI~4007. 
All calculations have been performed using the Helios computer cluster, 
which is an integral part of the MolDis 
Big Data facility, 
TIFR Hyderabad \href{http://moldis.tifrh.res.in}{(http://moldis.tifrh.res.in)}.
\end{acknowledgement}

\section*{Data Availability}
All data, analysis codes, RDKit-generated geometries, and DFT-optimized geometries associated with this work are available from the public repository at\\
\url{https://github.com/moldis-group/sears}.

\section*{Supporting Information}
Example code to retrieve 41 exothermic \(\mathrm{ROO^\bullet \rightarrow {}^\bullet QOOH}\) reactions from the SEARS dataset; 
Energy-profile diagrams for all 41 exothermic reactions. 

%\begin{suppinfo}

%\end{suppinfo}

\section*{Author Information}
\subsection*{Corresponding Author}
Raghunathan Ramakrishnan - Tata Institute of Fundamental Research, 36/P, Gopanpally Village, Serilingampally Mandal, Hyderabad, Telangana 500046, India; 
Orcid: \url{https://orcid.org/0000-0002-7288-9238}; 
Email: \texttt{ramakrishnan@tifrh.res.in}

\subsection*{Author contributions}
R.R. conceived the study, developed the workflow, performed all calculations and data analysis, and wrote the manuscript.

\subsection*{Notes}
The author declares no competing financial interest.

\bibliography{references}
% \bibliography{main-acs}

\end{document}

% \begin{tocentry}

% ...

% \end{tocentry}


\section*{Accessing the data}

The code below shows how the exothermic \(\mathrm{ROO^\bullet \rightarrow {}^\bullet QOOH}\) reactions can be identified from the validated SEARS dataset provided in the repository~\cite{sears2026github}. Entries are filtered according to the sign of \(E_{\mathrm{QOOH}} - E_{\mathrm{ROO}}\) using electronic energies calculated with $\omega$B97M-D4/def2-TZVP. 
Each retained reaction is represented by two rows corresponding to a diastereomeric transition-state pair. By choosing a value of \texttt{rxn\_index} between 0 and 40, the user can access any one of the 41 exothermic reactions and extract the corresponding transition-state energies relative to the \(\mathrm{R^\bullet + O_2}\) reference energy.

{
\footnotesize
\begin{verbatim}
import pandas as pd

E_O2 = -150.418813152134
au2kcm = 627.5095

df = pd.read_csv("../data-all/sears2324.csv").reset_index(drop=True)

pairs = []
for i in range(0, len(df), 2):
    row1 = df.iloc[i]
    row2 = df.iloc[i + 1]

    same_reaction = (
        row1["ROO_smiles"] == row2["ROO_smiles"]
        and row1["QOOH_smiles"] == row2["QOOH_smiles"]
    )
    if not same_reaction:
        raise ValueError(f"Rows {i} and {i+1} mismatch.")

    same_energies = (
        row1["ROO_Eele_TZVP"] == row2["ROO_Eele_TZVP"]
        and row1["QOOH_Eele_TZVP"] == row2["QOOH_Eele_TZVP"]
    )
    if not same_energies:
        raise ValueError(f"Rows {i} and {i+1} have inconsistent energies.")

    dE_rxn = (row1["QOOH_Eele_TZVP"] - row1["ROO_Eele_TZVP"]) * au2kcm
    if dE_rxn < 0:
        pairs.append(row1)
        pairs.append(row2)

df_filtered = pd.DataFrame(pairs).reset_index(drop=True)

print("Number of exothermic reactions:", len(df_filtered) // 2)

rxn_index = 0
row1 = df_filtered.iloc[2 * rxn_index]
row2 = df_filtered.iloc[2 * rxn_index + 1]

print("Reaction index:", rxn_index)
print("TS1:", (row1["TS_Eele_TZVP"] - row1["Rrad_Eele_TZVP"] - E_O2) * au2kcm)
print("TS2:", (row2["TS_Eele_TZVP"] - row2["Rrad_Eele_TZVP"] - E_O2) * au2kcm)
\end{verbatim}
}

The code returns the following output:
{
\footnotesize
\begin{verbatim}
Number of exothermic reactions: 41
Reaction index: 0
TS1: -18.813690633741153
TS2: -18.505989779122135
\end{verbatim}
}
\clearpage

\begin{figure}[p]
\centering
\includegraphics[width=0.75\linewidth]{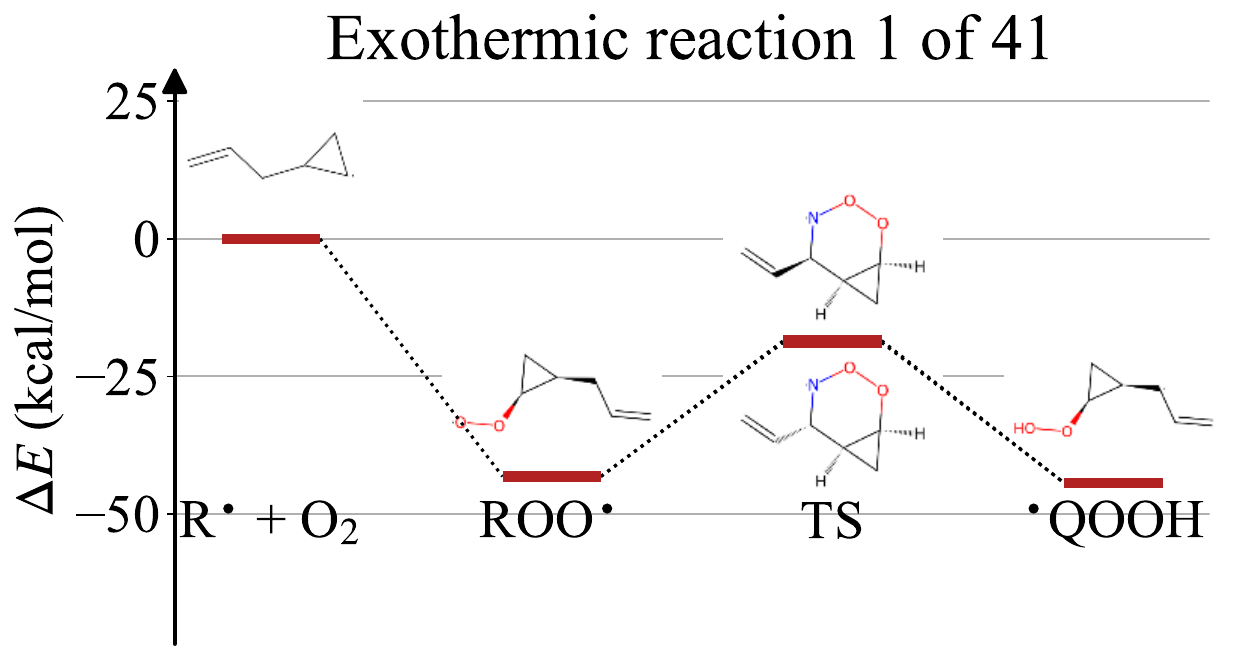}

\vspace{0.5cm}
\includegraphics[width=0.75\linewidth]{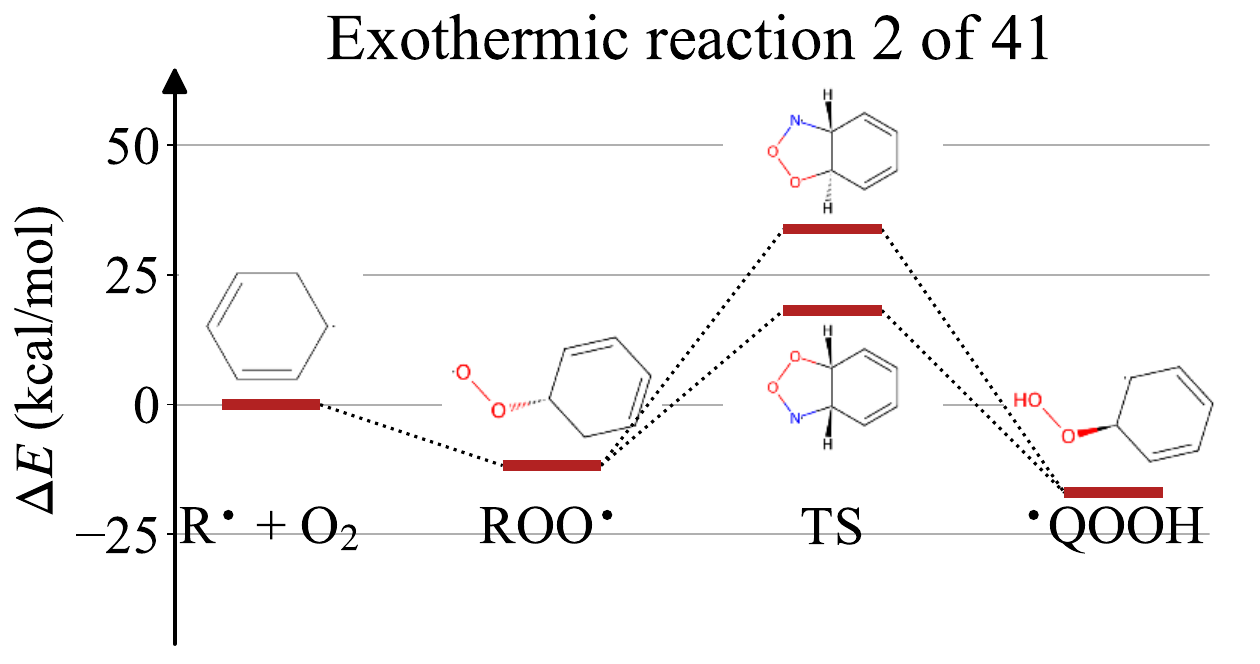}

\vspace{0.5cm}
\includegraphics[width=0.75\linewidth]{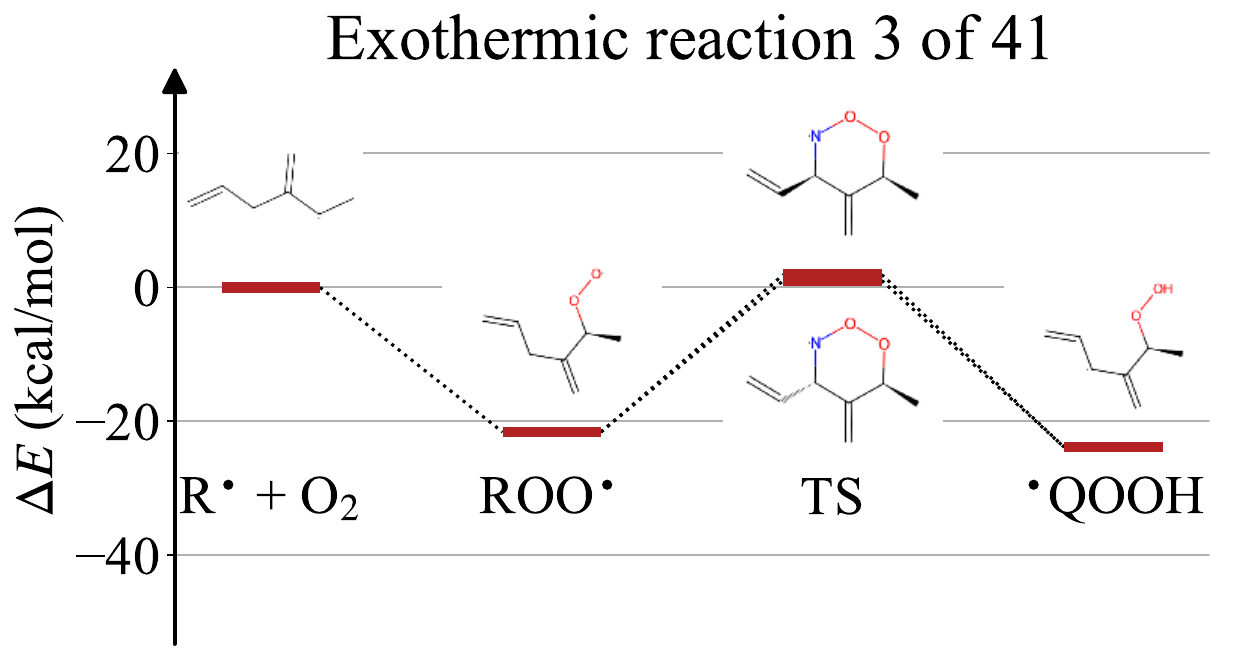}
\caption{
Representative energy profiles for selected diastereomeric \(\mathrm{ROO^{\bullet} \rightarrow {}^{\bullet}QOOH}\) isomerization reactions. Red horizontal lines denote stationary-point energies, dotted lines connect the reaction sequence, and molecular structures of the reactant, transition states, and product are shown for reference. 
N is used in the SMILES-based TS scaffold to represent the transferring H atom; all DFT calculations were performed after replacing N by H in the Cartesian coordinates.
}
\label{sifig:profiles}
\end{figure}

\begin{figure}[p]
\ContinuedFloat
\centering
\includegraphics[width=0.75\linewidth]{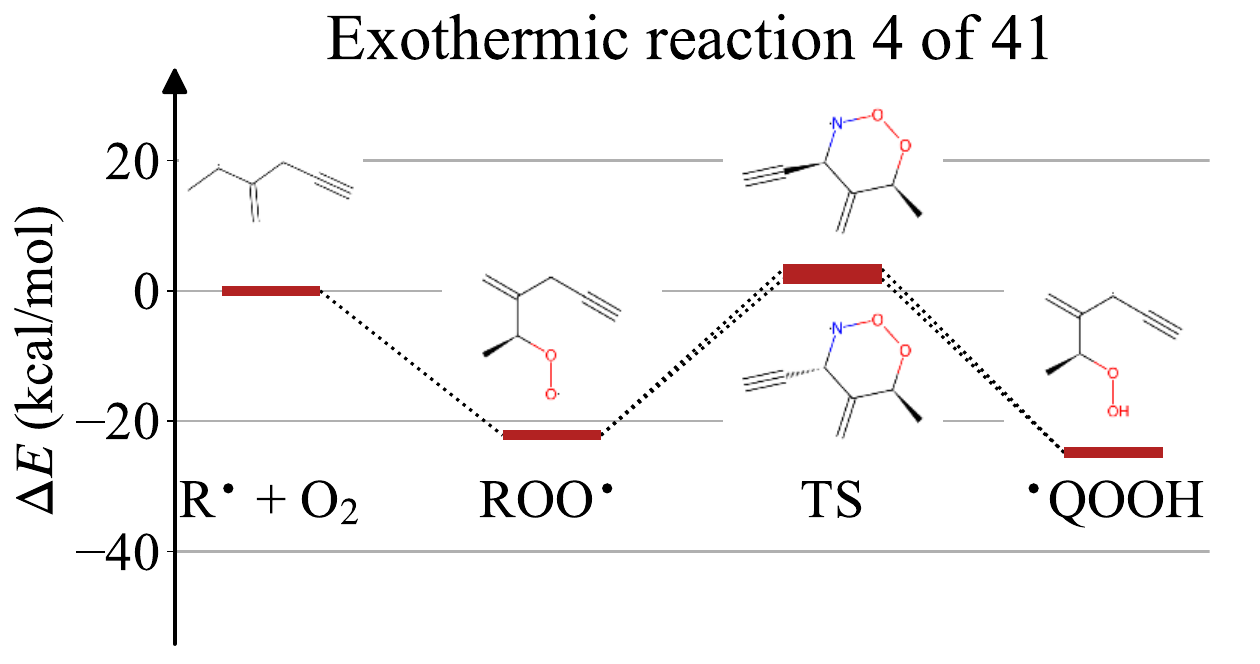}

\vspace{0.5cm}
\includegraphics[width=0.75\linewidth]{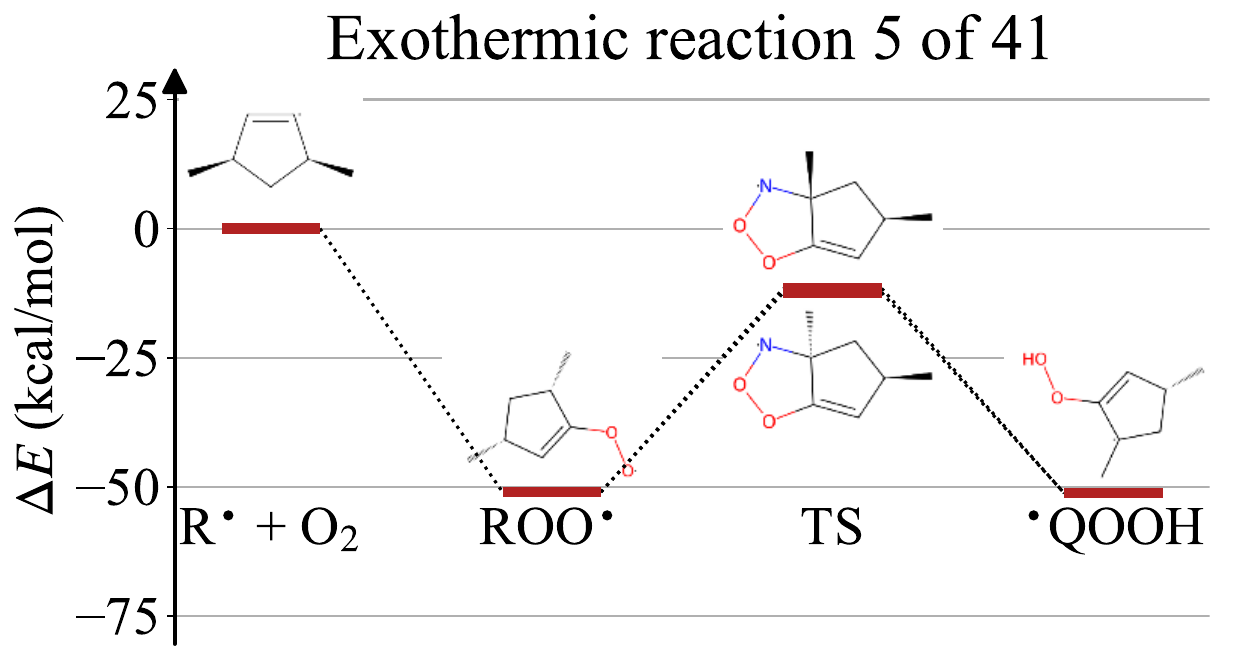}

\vspace{0.5cm}
\includegraphics[width=0.75\linewidth]{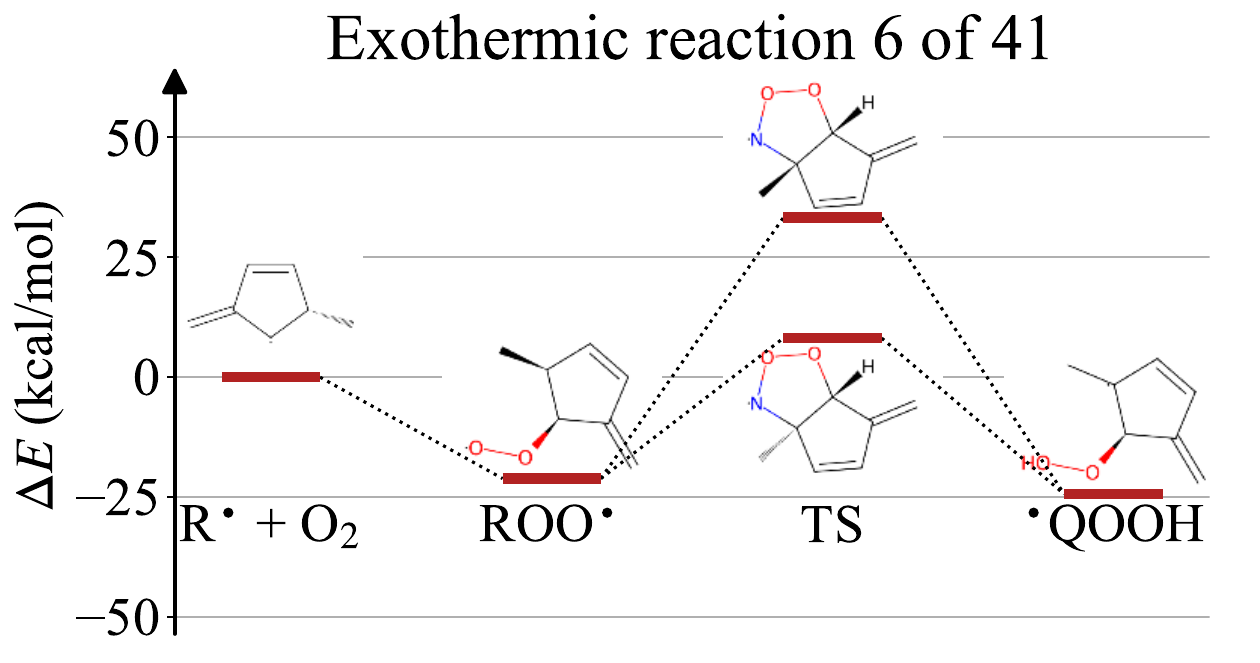}
\caption[]{(continued from previous page).}
\label{sifig:profiles_cont}
\end{figure}

\begin{figure}[p]
\ContinuedFloat
\centering
\includegraphics[width=0.75\linewidth]{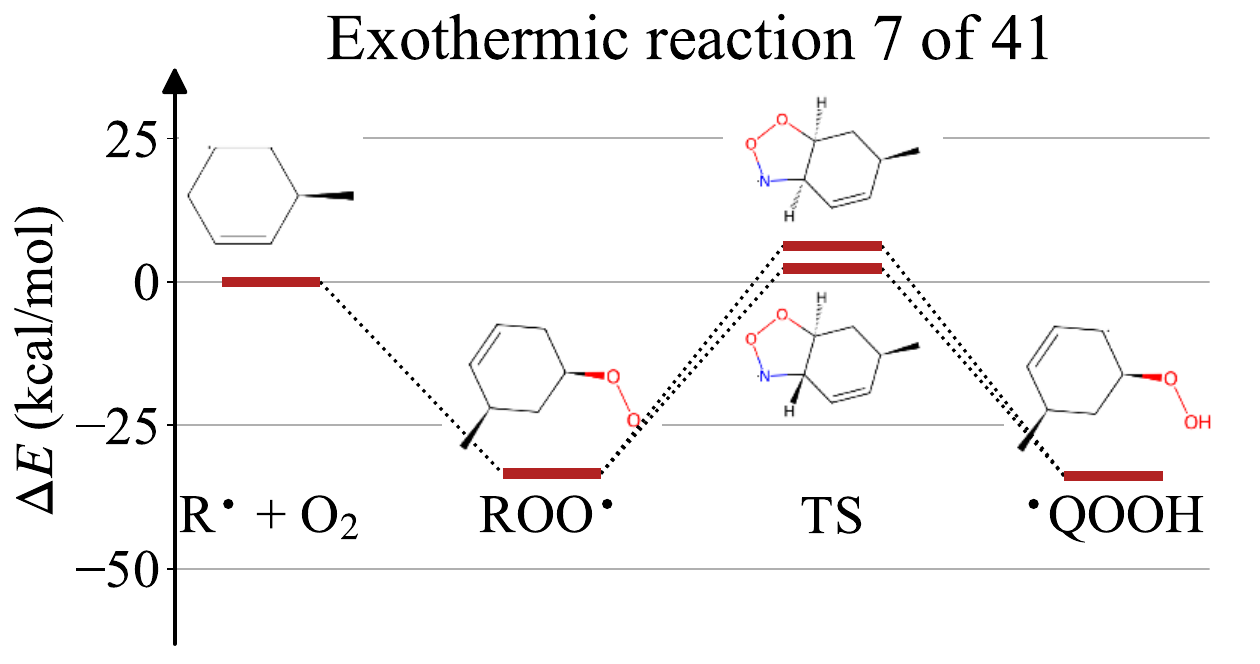}

\vspace{0.5cm}
\includegraphics[width=0.75\linewidth]{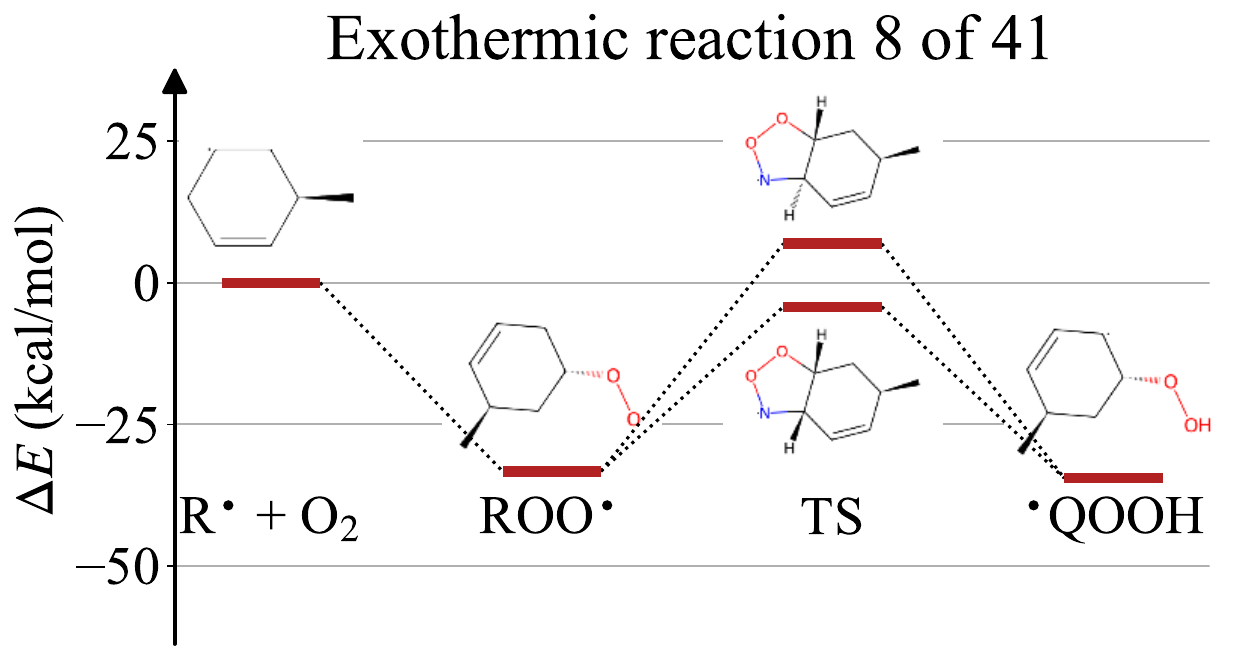}

\vspace{0.5cm}
\includegraphics[width=0.75\linewidth]{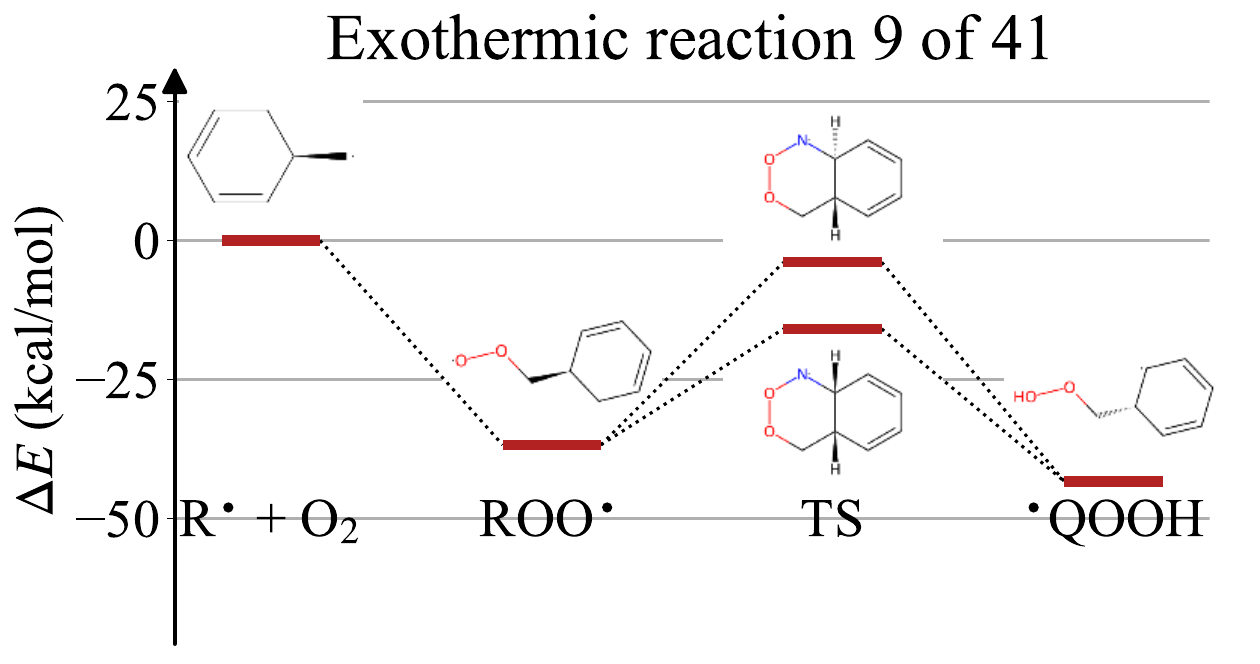}
\caption[]{(continued from previous page).}
\label{sifig:profiles_cont}
\end{figure}

\begin{figure}[p]
\ContinuedFloat
\centering
\includegraphics[width=0.75\linewidth]{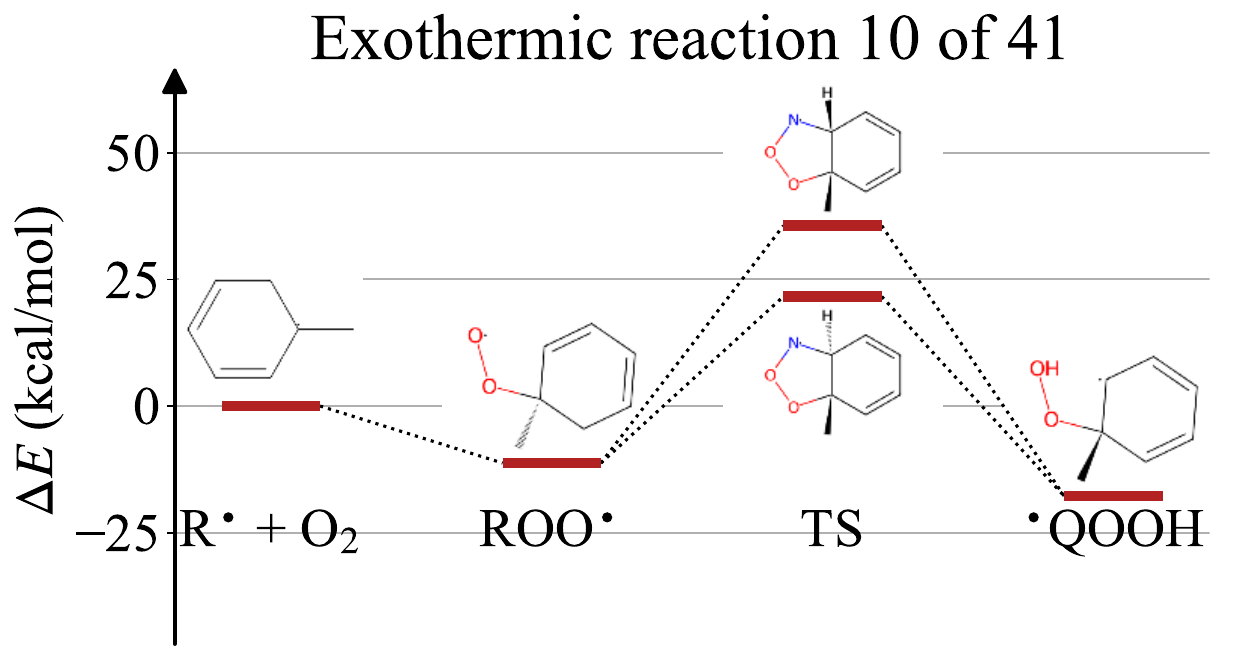}

\vspace{0.5cm}
\includegraphics[width=0.75\linewidth]{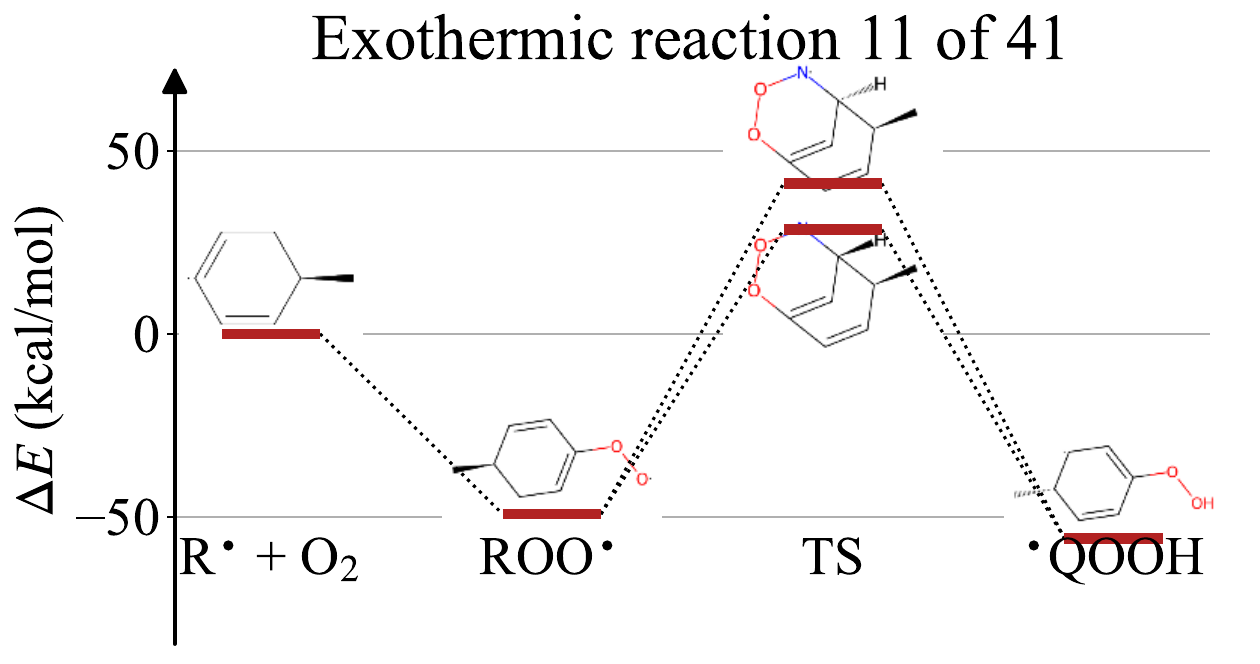}

\vspace{0.5cm}
\includegraphics[width=0.75\linewidth]{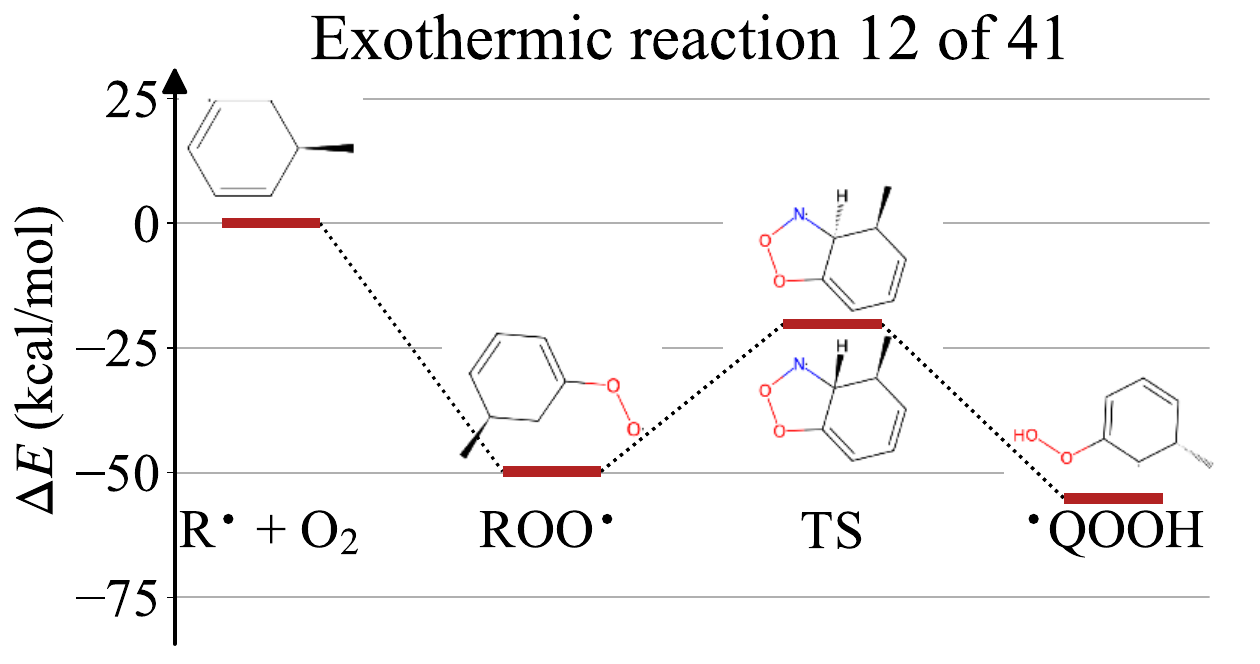}
\caption[]{(continued from previous page).}
\label{sifig:profiles_cont_9}
\end{figure}

\begin{figure}[p]
\ContinuedFloat
\centering
\includegraphics[width=0.75\linewidth]{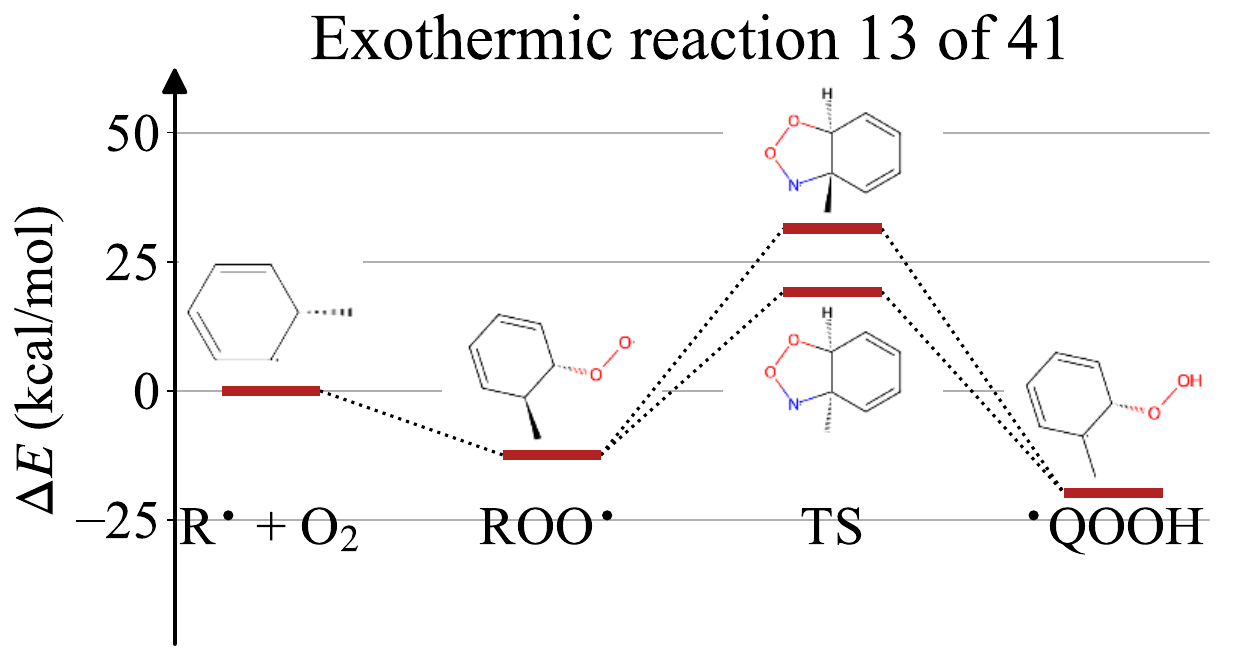}

\vspace{0.5cm}
\includegraphics[width=0.75\linewidth]{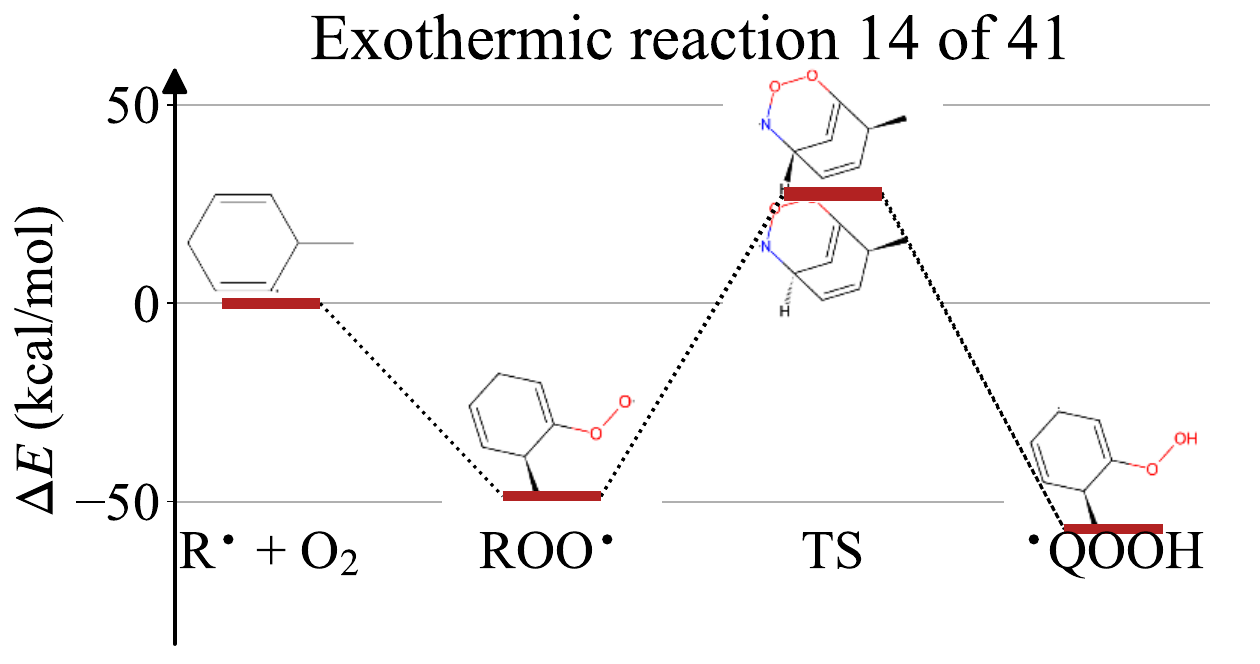}

\vspace{0.5cm}
\includegraphics[width=0.75\linewidth]{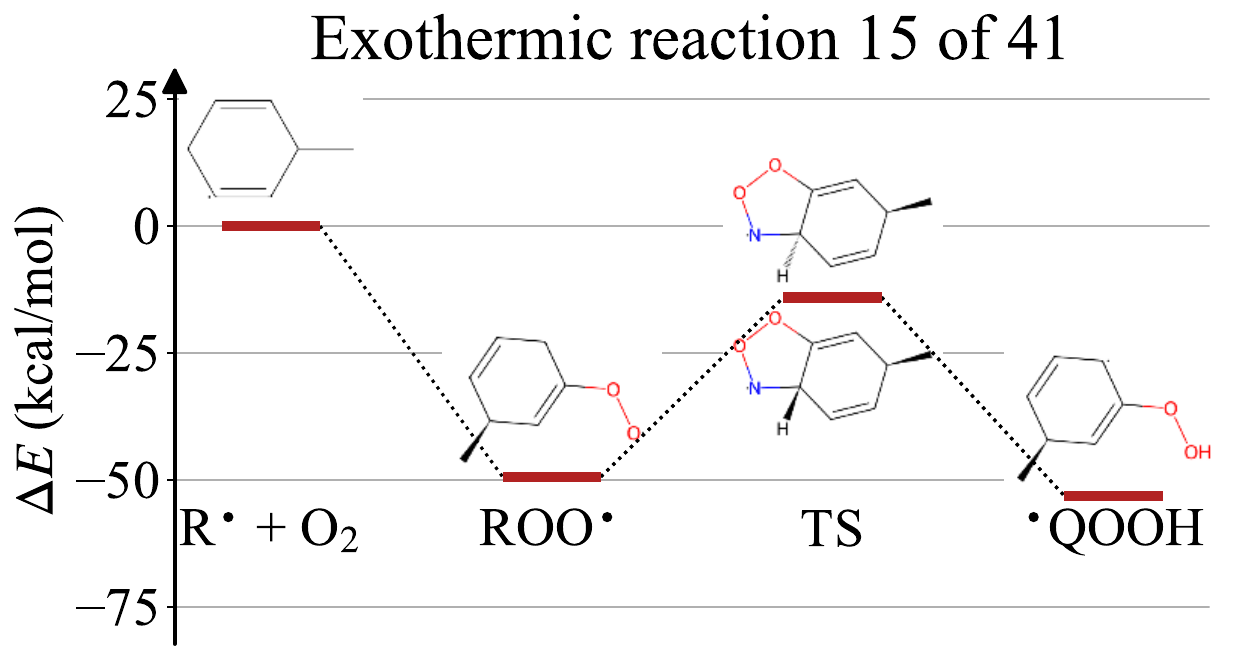}
\caption[]{(continued from previous page).}
\label{sifig:profiles_cont_12}
\end{figure}

\begin{figure}[p]
\ContinuedFloat
\centering
\includegraphics[width=0.75\linewidth]{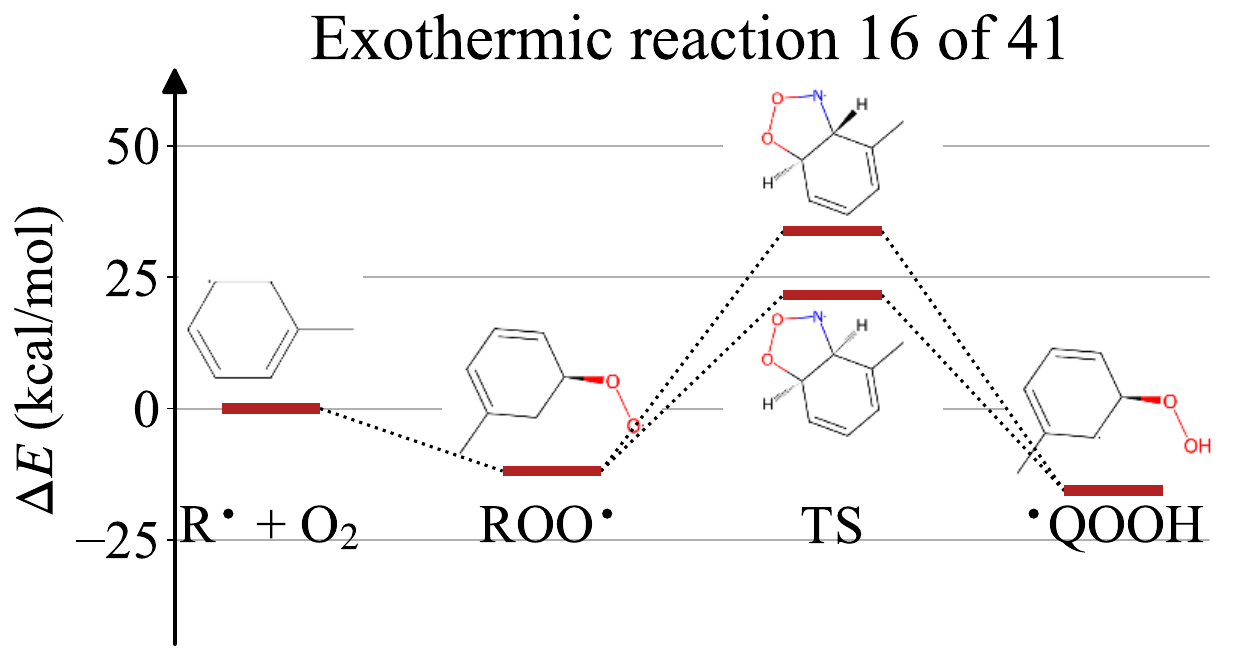}

\vspace{0.5cm}
\includegraphics[width=0.75\linewidth]{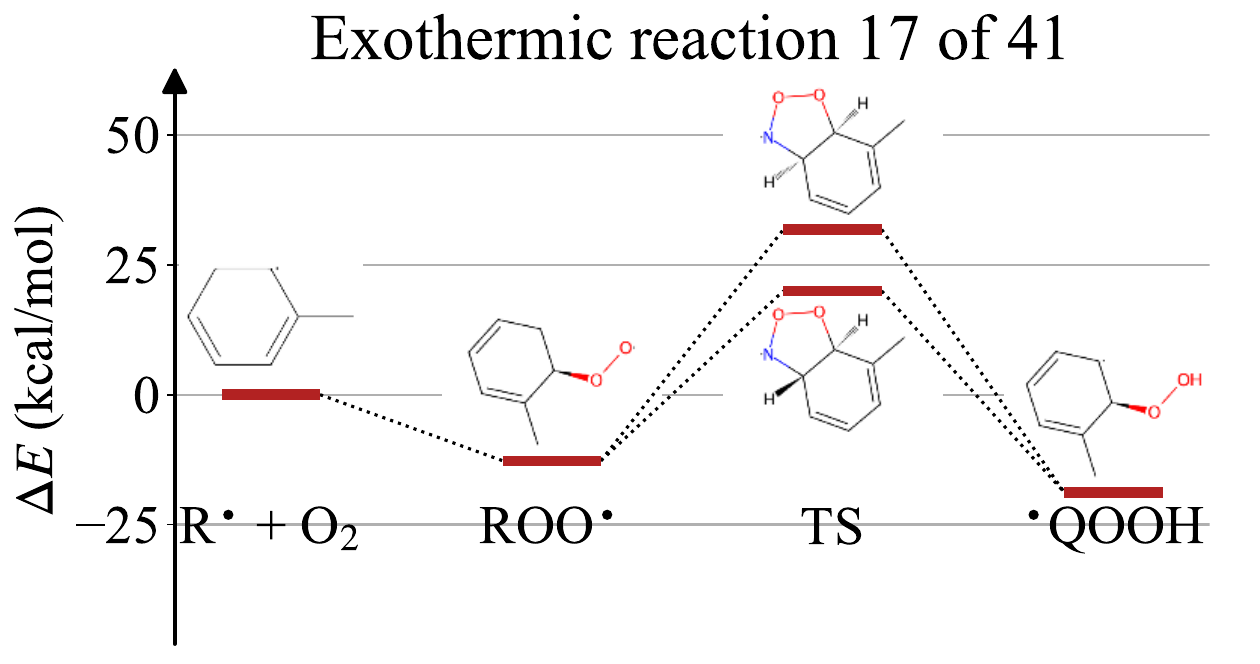}

\vspace{0.5cm}
\includegraphics[width=0.75\linewidth]{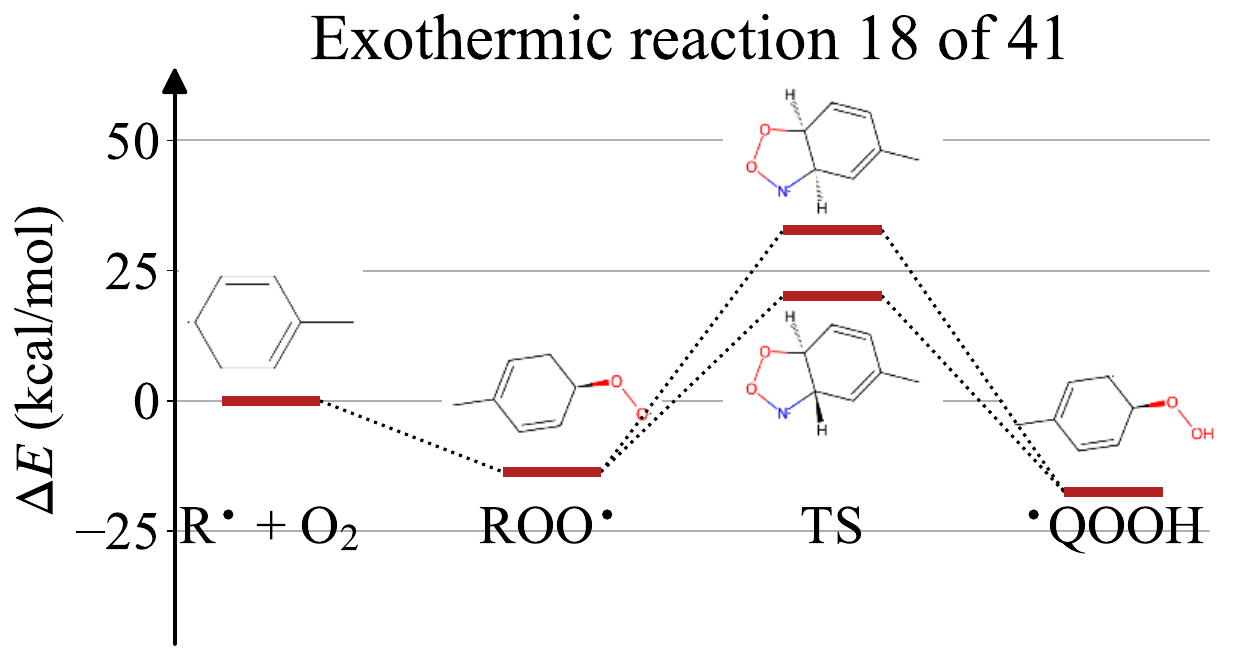}
\caption[]{(continued from previous page).}
\label{sifig:profiles_cont_15}
\end{figure}

\begin{figure}[p]
\ContinuedFloat
\centering
\includegraphics[width=0.75\linewidth]{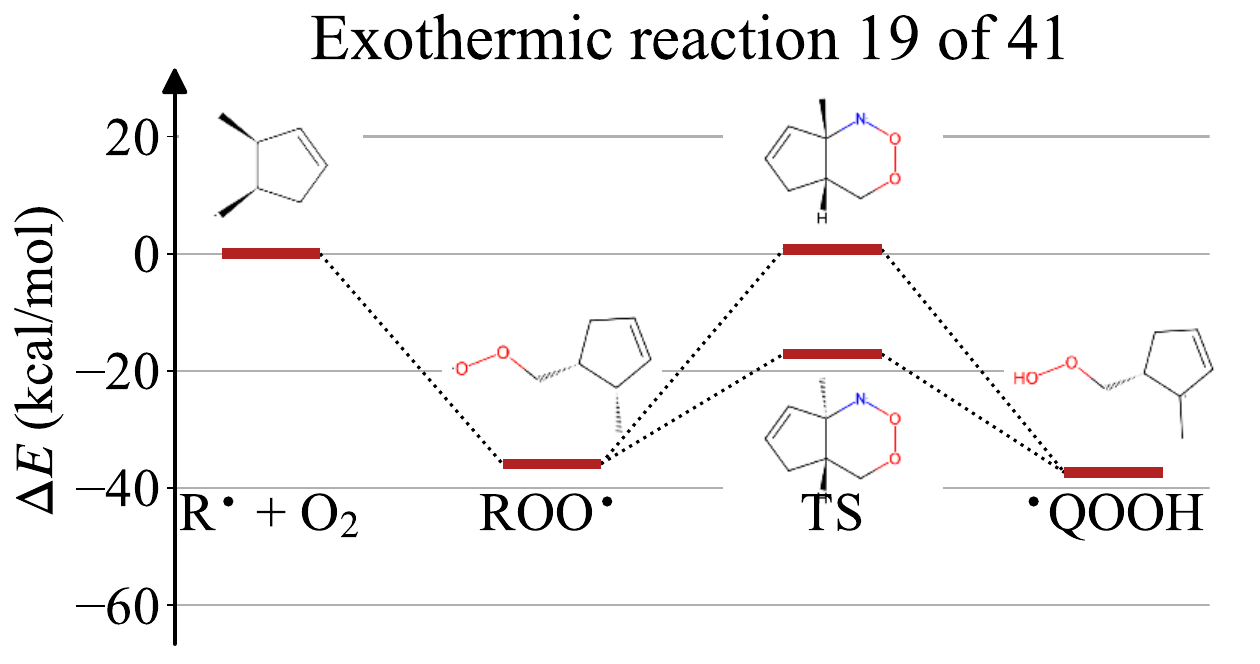}

\vspace{0.5cm}
\includegraphics[width=0.75\linewidth]{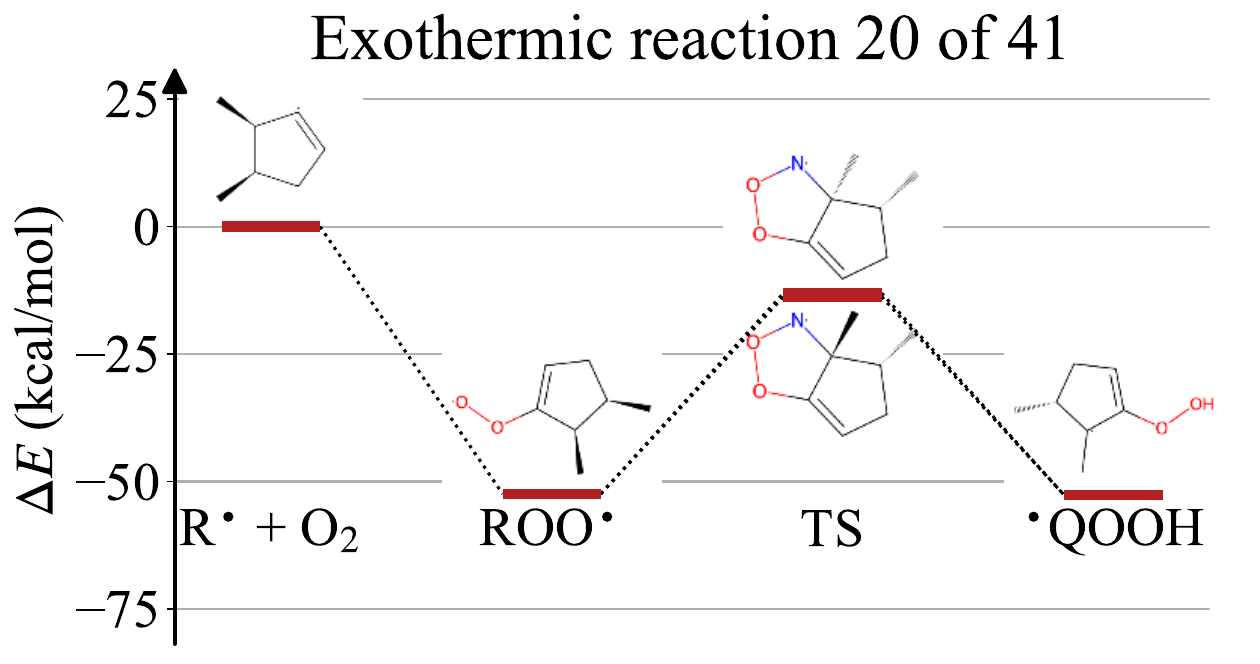}

\vspace{0.5cm}
\includegraphics[width=0.75\linewidth]{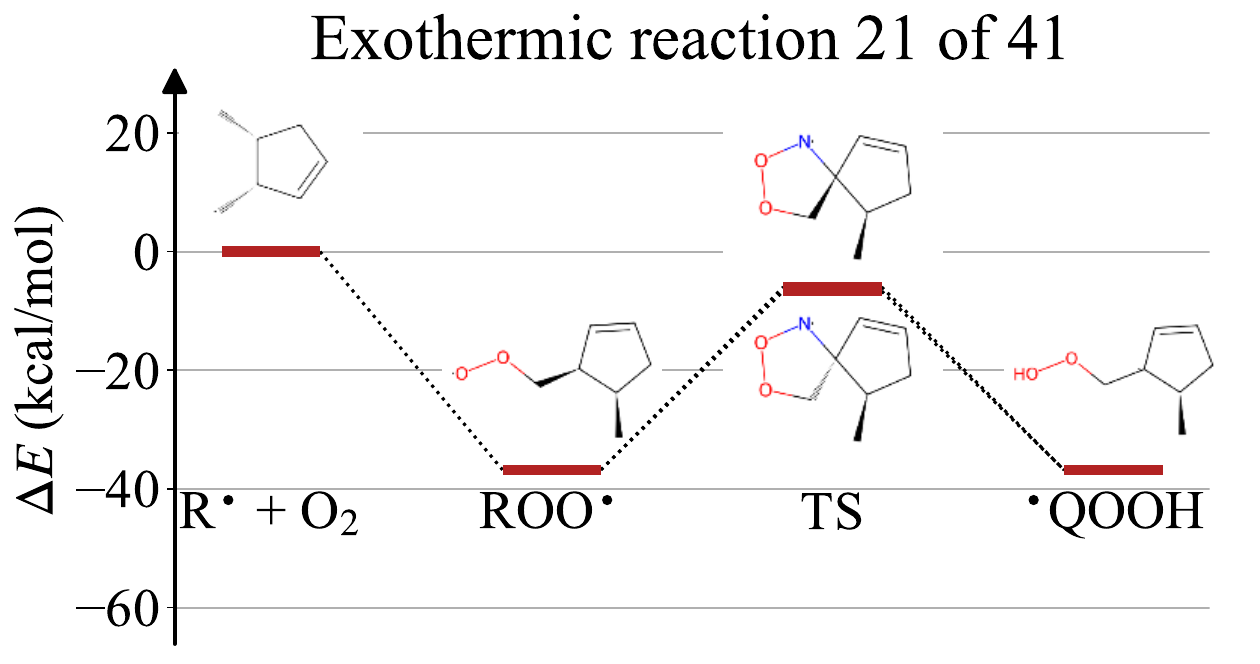}
\caption[]{(continued from previous page).}
\label{sifig:profiles_cont_18}
\end{figure}

\begin{figure}[p]
\ContinuedFloat
\centering
\includegraphics[width=0.75\linewidth]{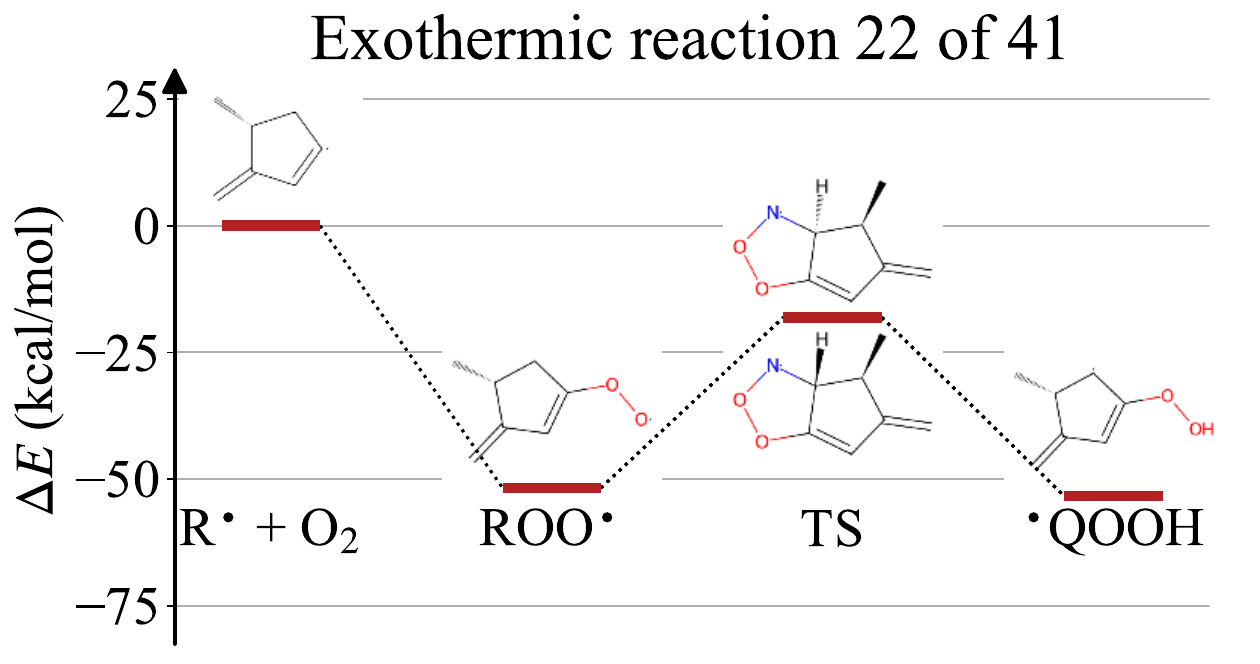}

\vspace{0.5cm}
\includegraphics[width=0.75\linewidth]{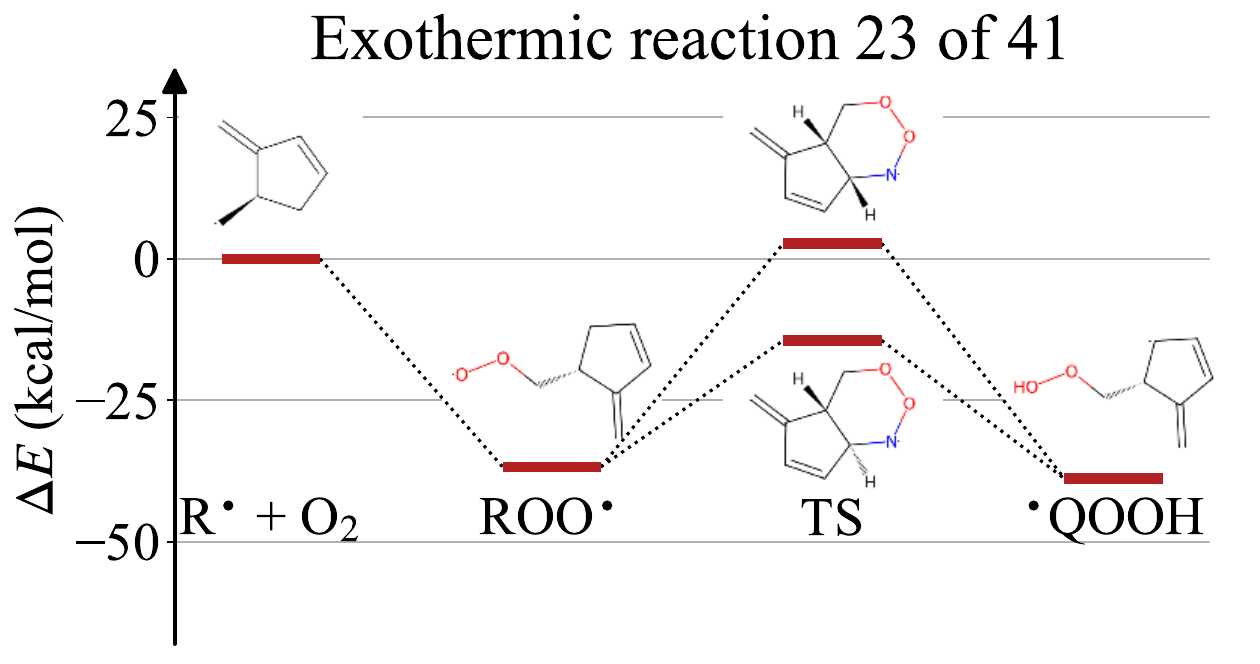}

\vspace{0.5cm}
\includegraphics[width=0.75\linewidth]{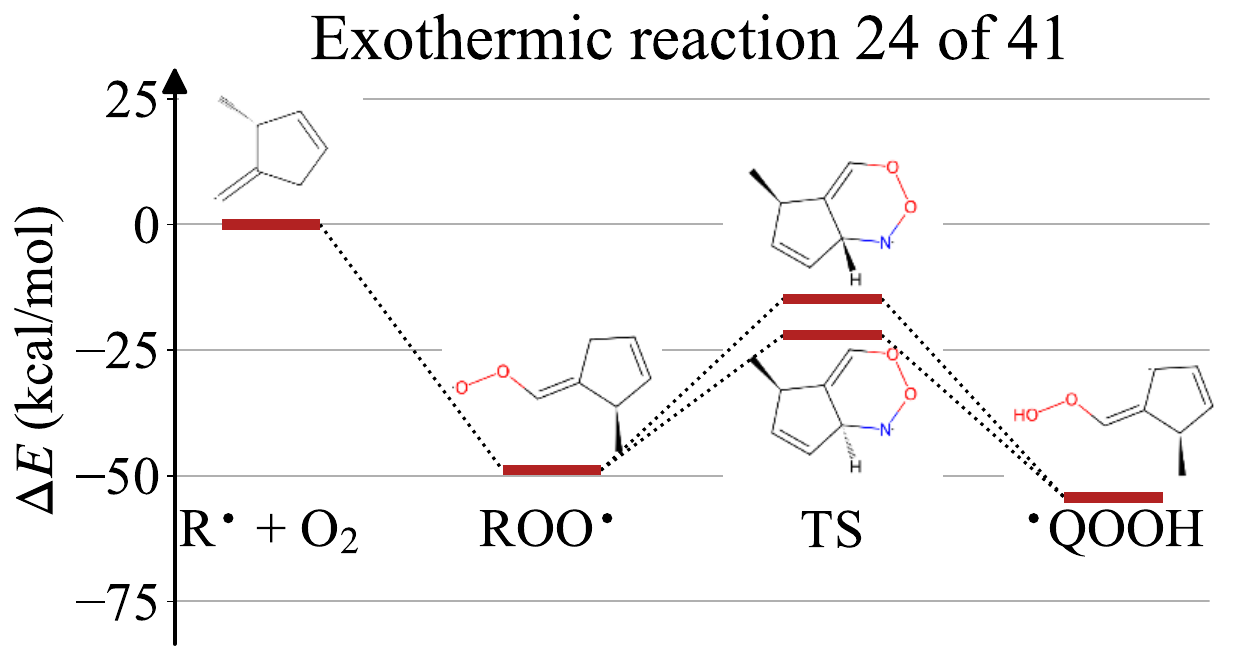}
\caption[]{(continued from previous page).}
\label{sifig:profiles_cont_21}
\end{figure}

\begin{figure}[p]
\ContinuedFloat
\centering
\includegraphics[width=0.75\linewidth]{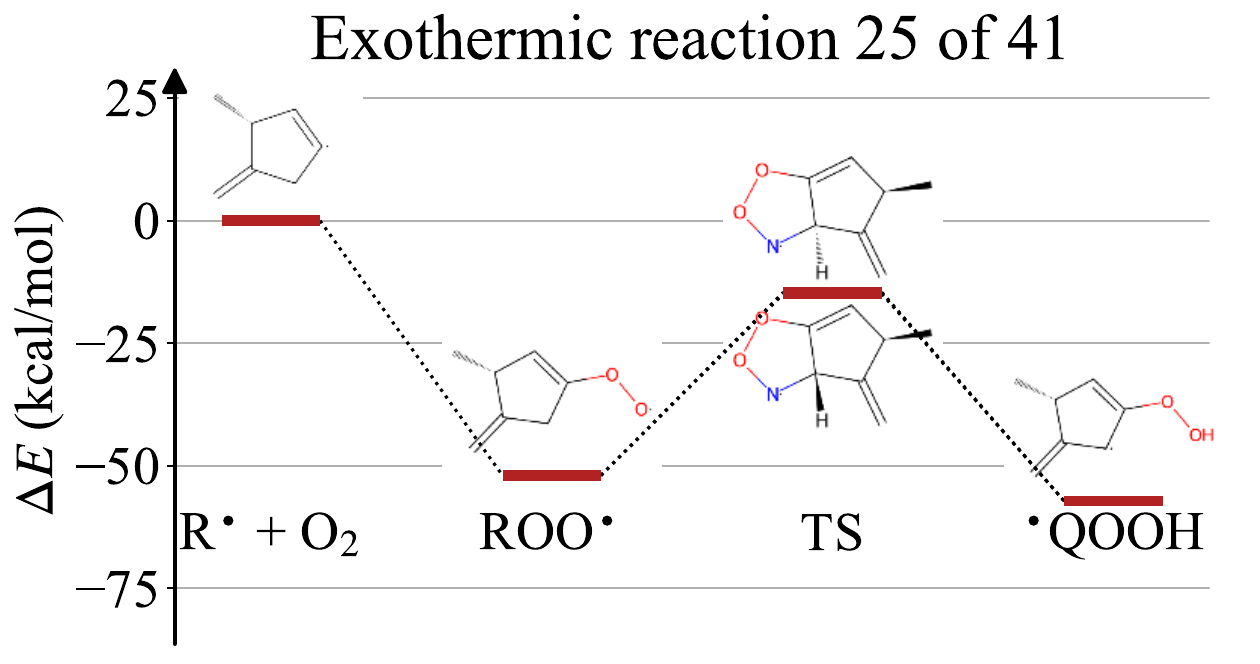}

\vspace{0.5cm}
\includegraphics[width=0.75\linewidth]{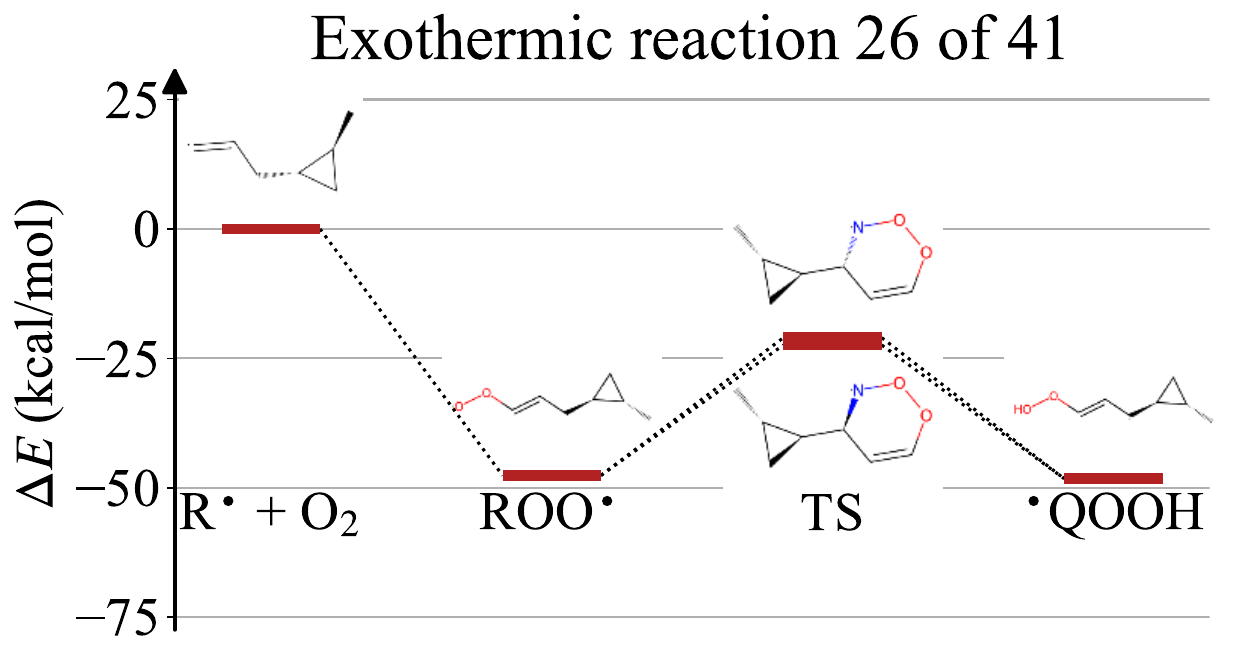}

\vspace{0.5cm}
\includegraphics[width=0.75\linewidth]{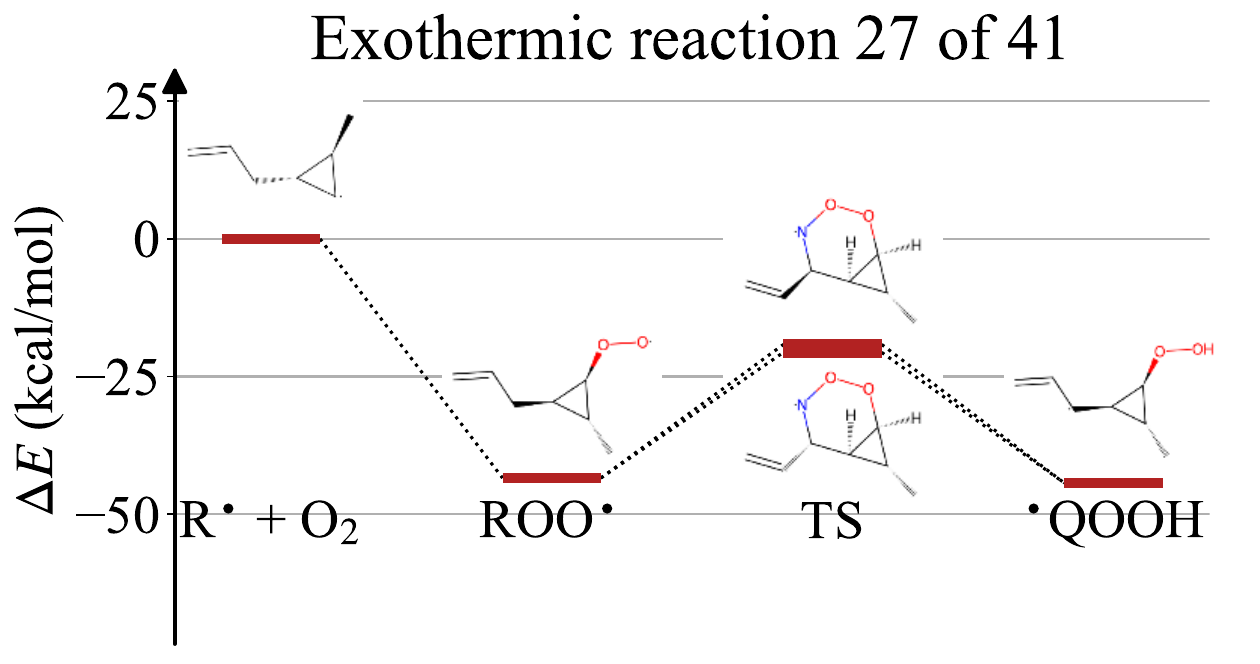}
\caption[]{(continued from previous page).}
\label{sifig:profiles_cont_24}
\end{figure}

\begin{figure}[p]
\ContinuedFloat
\centering
\includegraphics[width=0.75\linewidth]{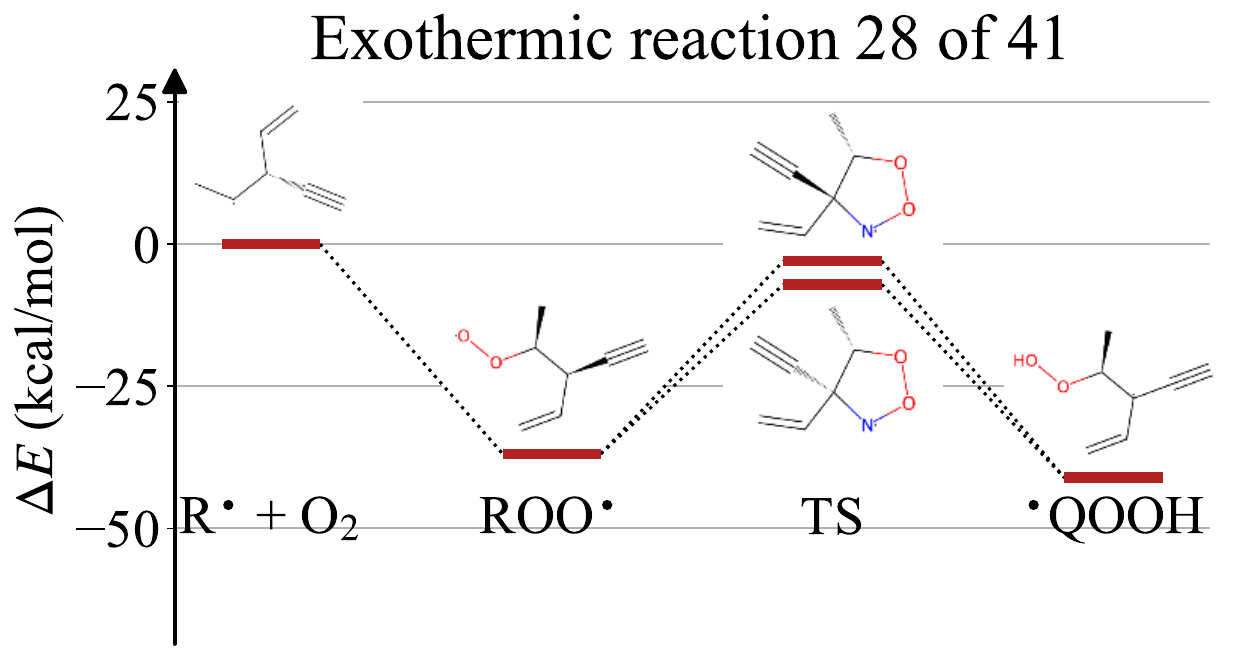}

\vspace{0.5cm}
\includegraphics[width=0.75\linewidth]{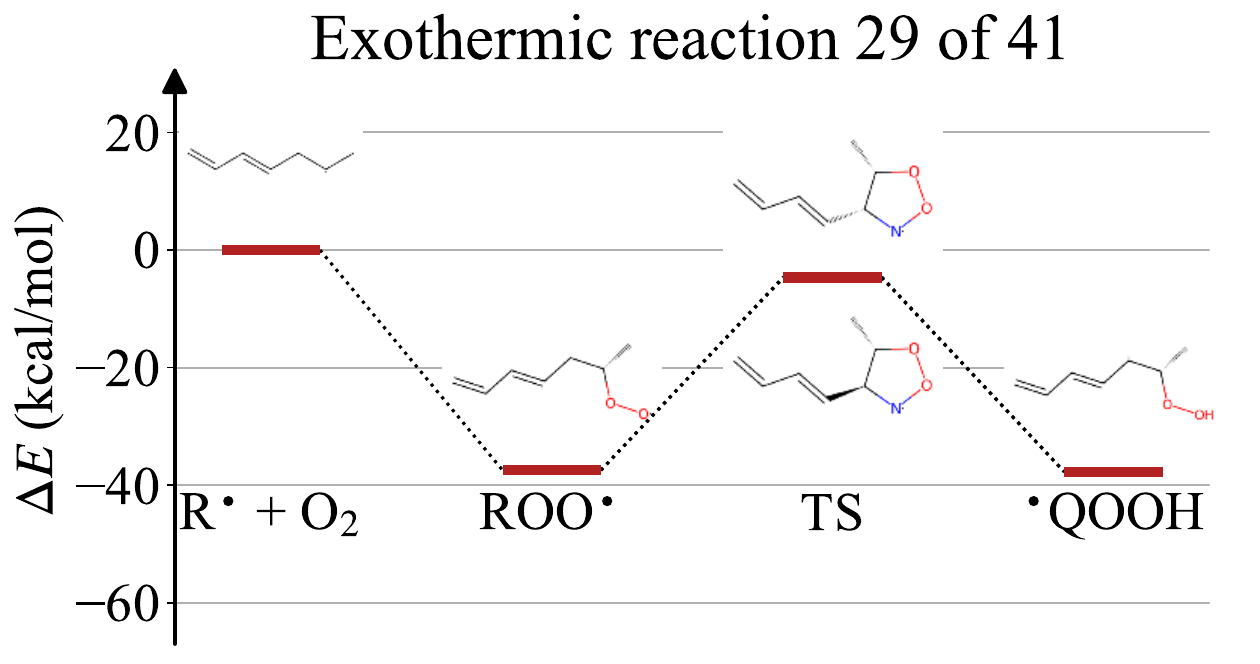}

\vspace{0.5cm}
\includegraphics[width=0.75\linewidth]{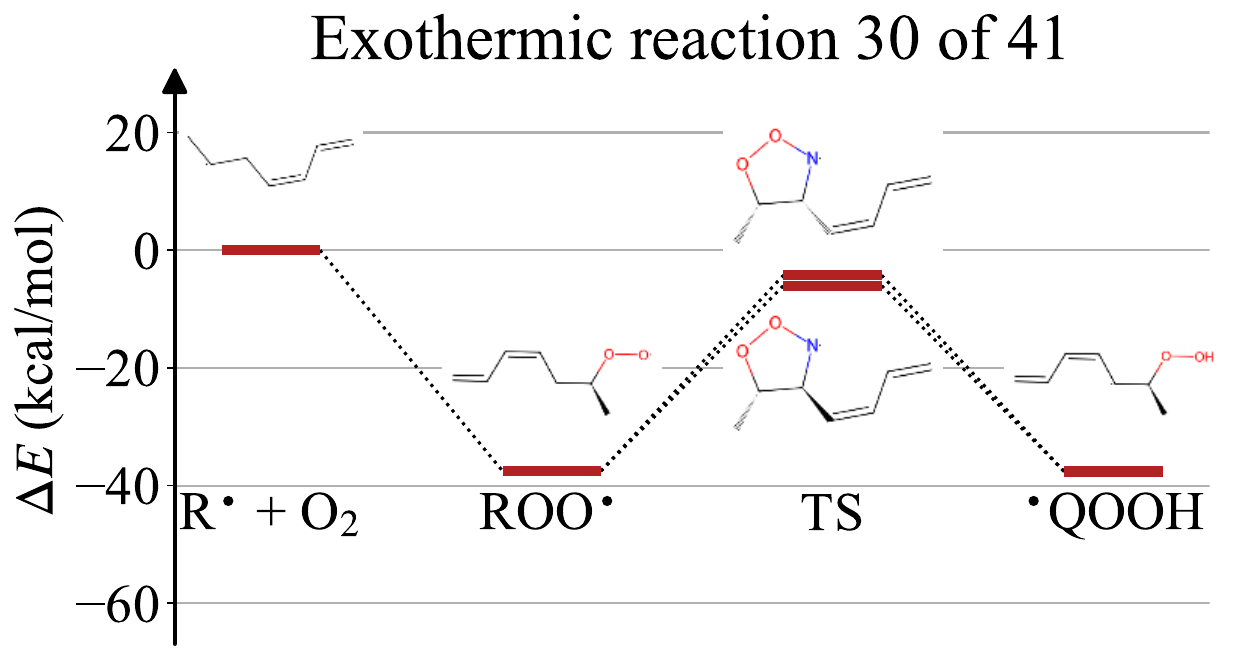}
\caption[]{(continued from previous page).}
\label{sifig:profiles_cont_27}
\end{figure}

\begin{figure}[p]
\ContinuedFloat
\centering
\includegraphics[width=0.75\linewidth]{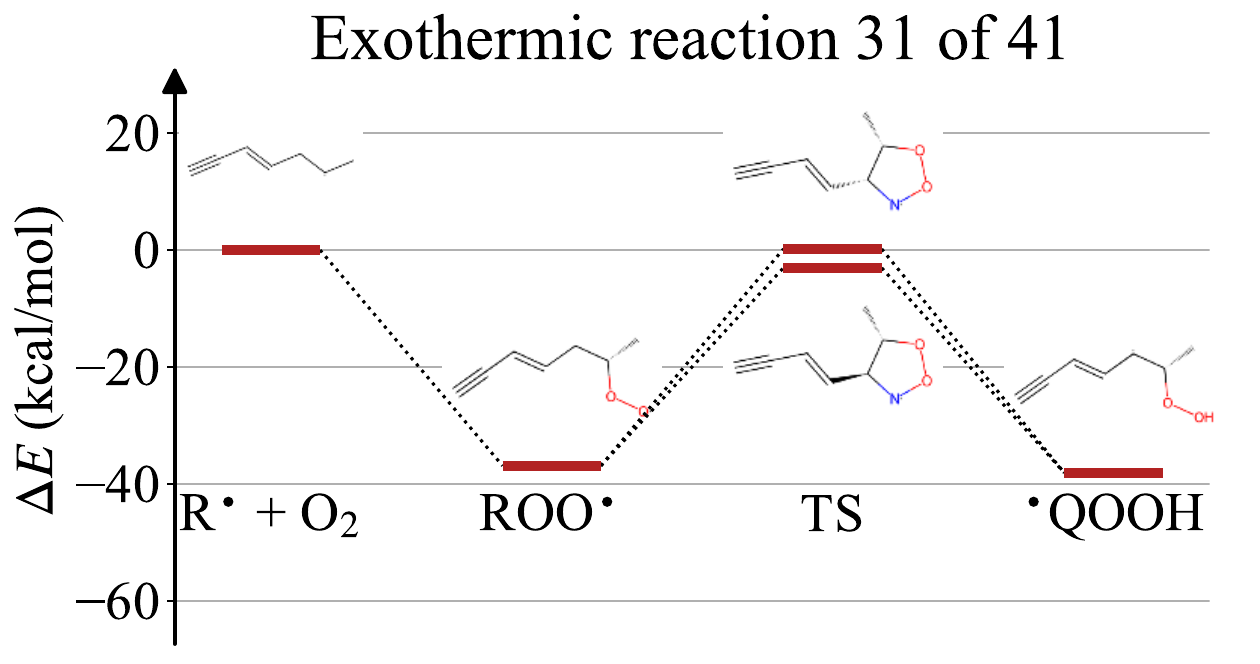}

\vspace{0.5cm}
\includegraphics[width=0.75\linewidth]{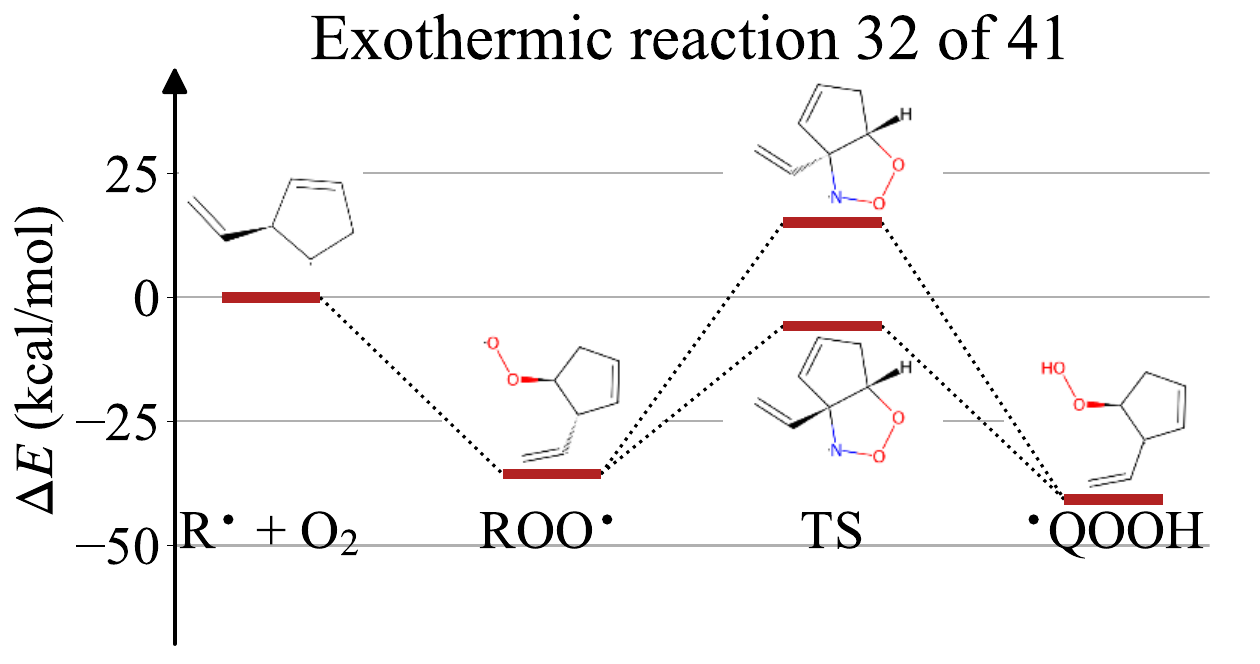}

\vspace{0.5cm}
\includegraphics[width=0.75\linewidth]{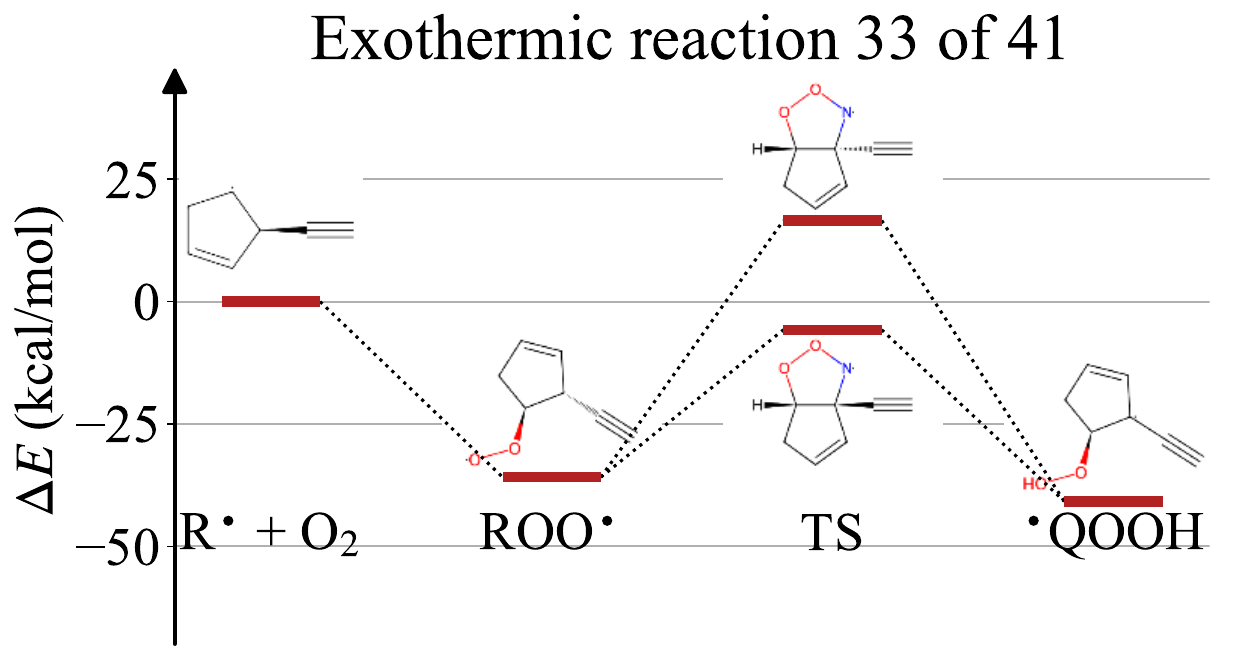}
\caption[]{(continued from previous page).}
\label{sifig:profiles_cont_30}
\end{figure}

\begin{figure}[p]
\ContinuedFloat
\centering
\includegraphics[width=0.75\linewidth]{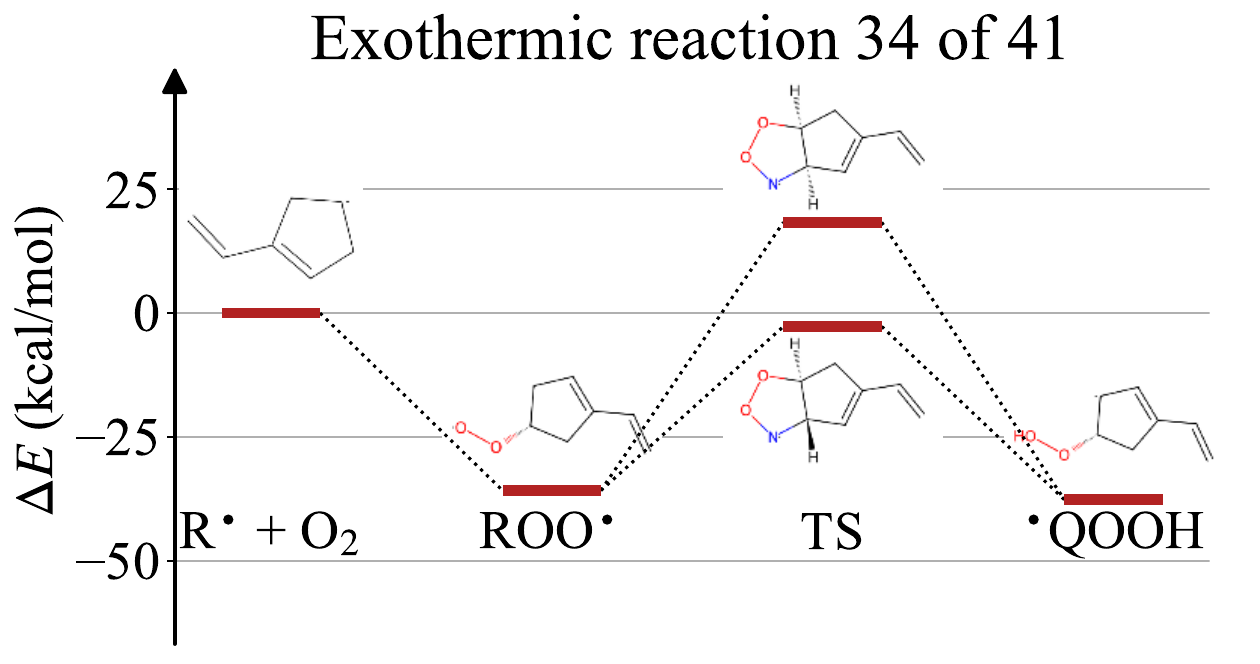}

\vspace{0.5cm}
\includegraphics[width=0.75\linewidth]{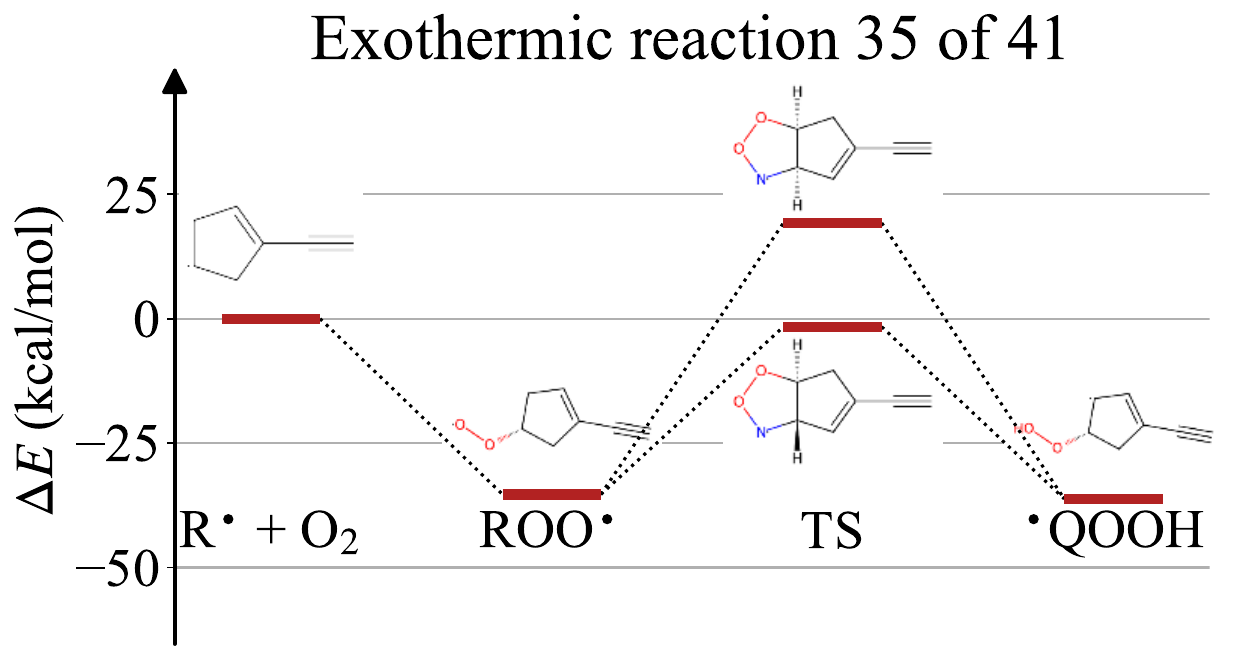}

\vspace{0.5cm}
\includegraphics[width=0.75\linewidth]{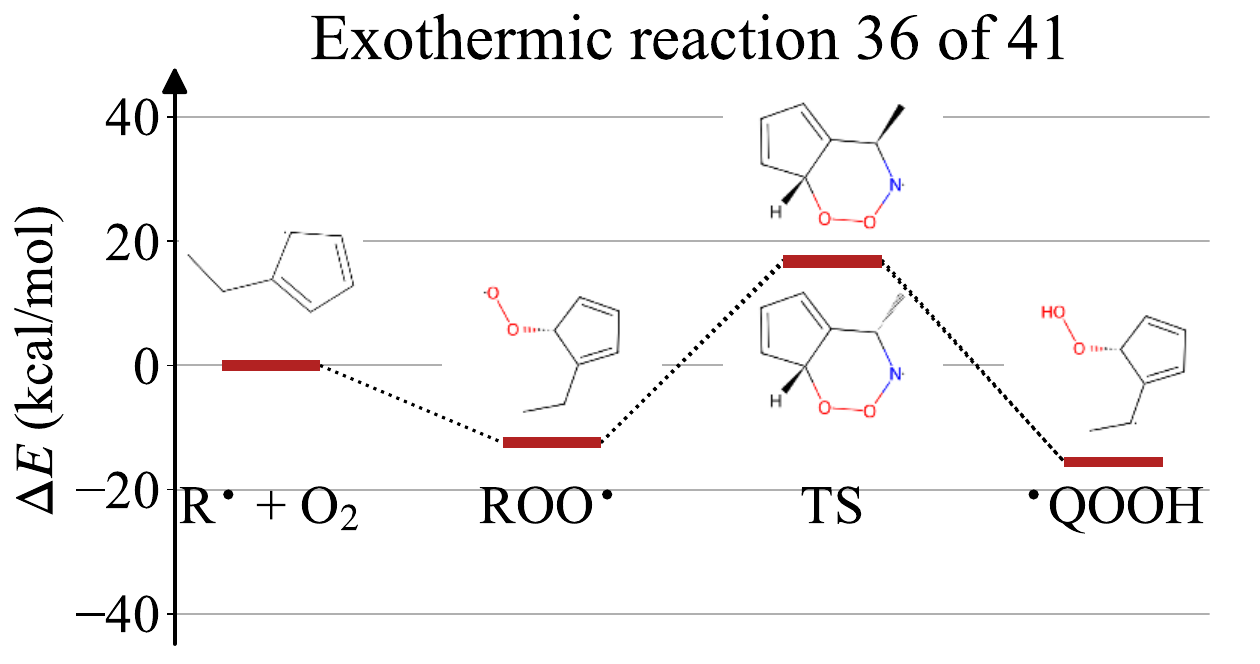}
\caption[]{(continued from previous page).}
\label{sifig:profiles_cont_33}
\end{figure}

\begin{figure}[p]
\ContinuedFloat
\centering
\includegraphics[width=0.75\linewidth]{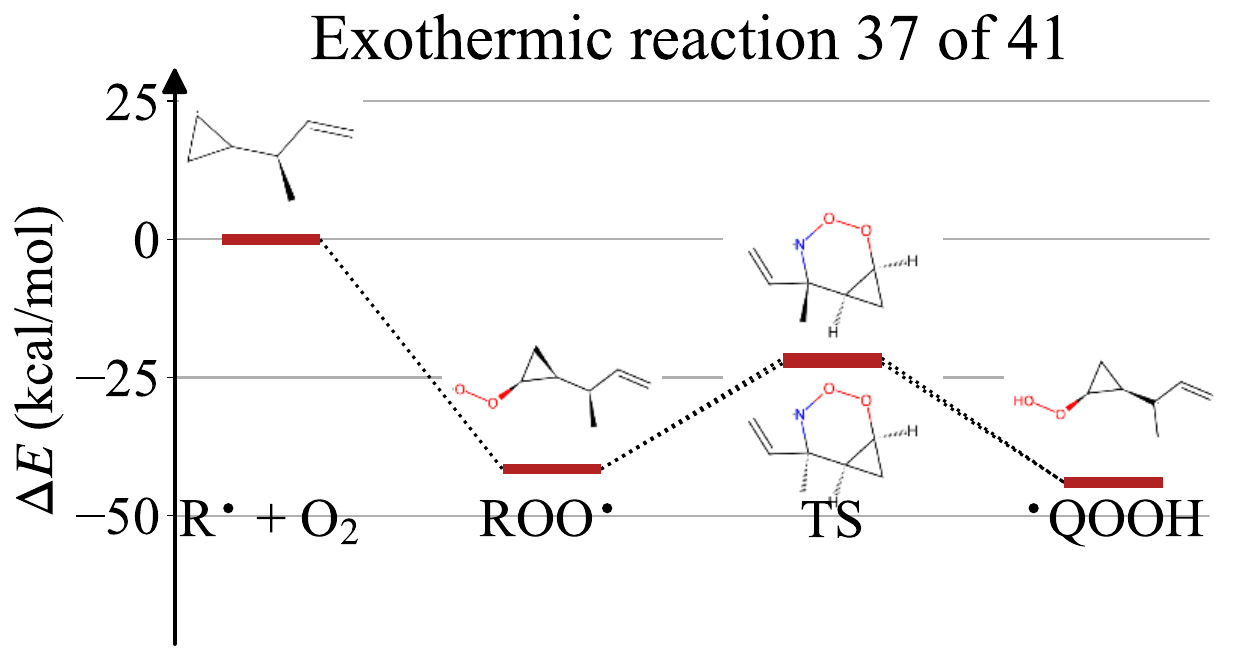}

\vspace{0.5cm}
\includegraphics[width=0.75\linewidth]{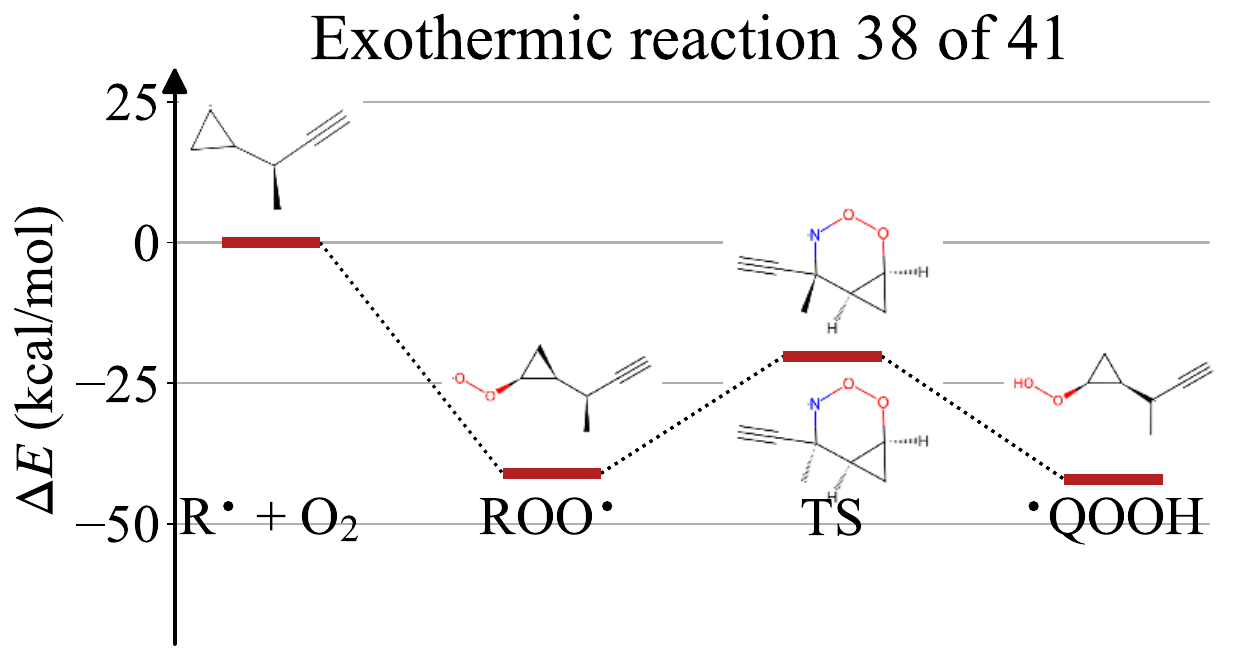}

\vspace{0.5cm}
\includegraphics[width=0.75\linewidth]{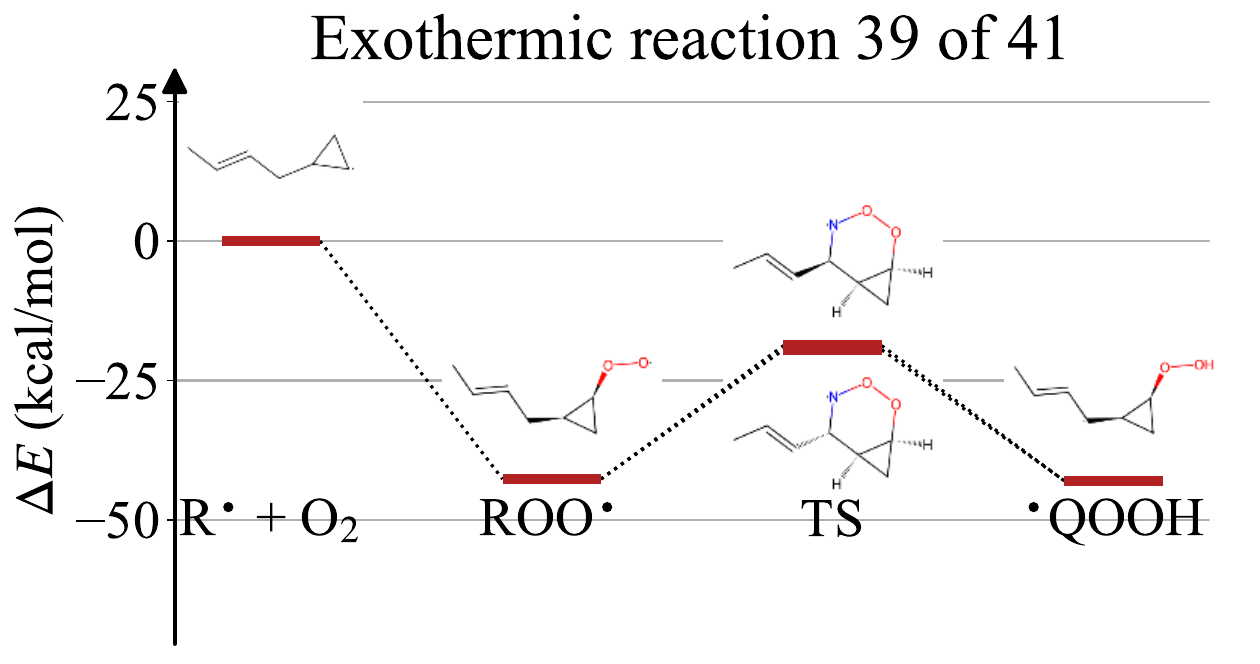}
\caption[]{(continued from previous page).}
\label{sifig:profiles_cont_36}
\end{figure}

\begin{figure}[p]
\ContinuedFloat
\centering
\includegraphics[width=0.75\linewidth]{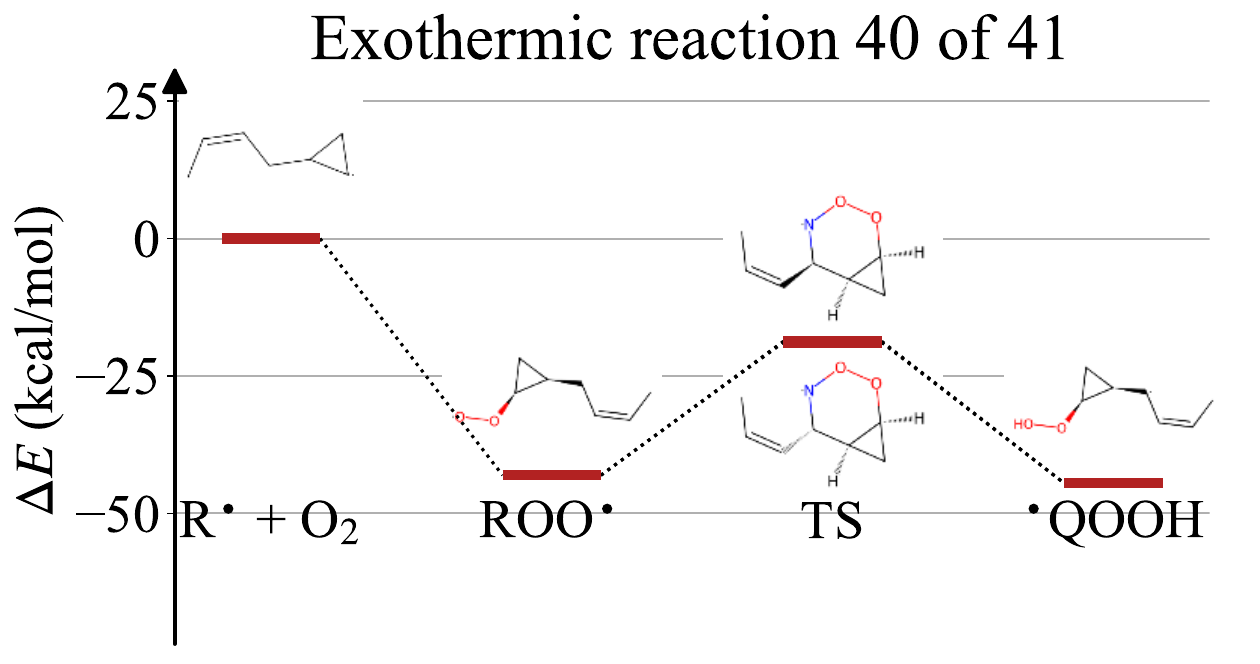}

\vspace{0.5cm}
\includegraphics[width=0.75\linewidth]{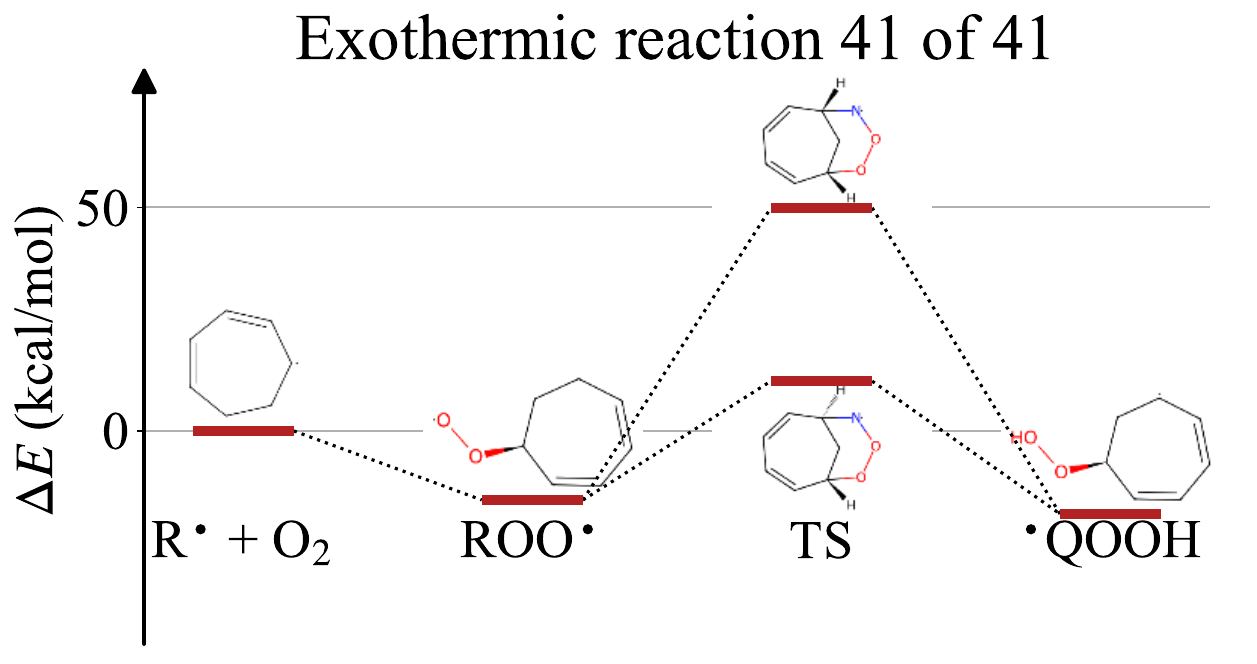}
\caption[]{(continued from previous page).}
\label{sifig:profiles_cont_39}
\end{figure}


\clearpage
\bibliography{references}